\documentstyle[aps,12pt]{revtex}
\begin{document}
\draft
\title{\Large \bf Filling transition for a wedge
\vspace*{5mm}}
\author{{\large K. Rejmer $^{1,2}$, S. Dietrich $^1$, 
and M. Napi\'orkowski $^2$}\\
\it $^{1}$ Fachbereich Physik, Bergische Universit\"{a}t
Wuppertal,\\
\it D-42097 Wuppertal, Germany\\
\it $^{2}$ Instytut Fizyki Teoretycznej, Uniwersytet 
Warszawski,\\
\it 00-681 Warszawa, Ho\.za 69, Poland
\vspace*{5mm}}
\maketitle

\centerline{\bf Abstract}

\begin{abstract}

We study the formation and the shape of a liquid meniscus in a
wedge with opening angle $2\,\varphi$ which is exposed to a vapor
phase. By applying a suitable effective interface model, at
liquid-vapor coexistence and at a temperature $T_{\varphi}$
we find a filling transition at which the height of the meniscus
becomes macroscopically large while the planar walls of the wedge
far away from its center remain nonwet up to the wetting transition
occurring at $T_w>T_{\varphi}$. Depending on the fluid and the
substrate potential the filling transition can be either continuous
or discontinuous. In the latter case it is accompanied by a
owprefilling line extending into the vapor phase of the bulk phase
diagram and describing a transition from a small to a large, but
finite, meniscus height. The filling and the prefilling transitions
correspond to nonanalyticities in the surface and line contributions to
the free energy of the fluid, respectively.
\end{abstract}

\newpage

\centerline{\bf {I. Introduction}}
\renewcommand{\theequation}{1.\arabic{equation}} 

Unless much experimental care is provided in growing crystals and
cutting them, genericly on an atomic scale their surfaces exhibit
an irregular topography. Numerous theoretical [1-10] and experimental 
[11-14] 
studies have demonstrated that if such real surfaces are exposed to
a vapor phase the resulting wetting phenomena [15,16] may differ
significantly from the corresponding ones on perfectly planar surfaces
of the same substrate-fluid systems. These studies are focused on
the properties of the adsorbed liquid films averaged laterally over
the statistical irregularities of the substrate topography.

However, e.g., in the context of structured semiconductor
surfaces [17-19], microfluidics [20-24], and templates for the selfassembly of
small particles [25-27] highly {\it regular} nonflat lateral surface structures
can be produced. But both the theoretical understanding of the
{\it local} wetting phenomena in such structures [28-42] as well as
corresponding experimental investigations [43,44] are still in their
infancies [45].

As a paradigm of such structures we consider a single wedgelike cavity
with an opening angle $2\,{\varphi}$ and macroscopic extension along
its edge; along this $y$ direction the system is taken to be 
translationally invariant (see Fig.1(a)). The wedge is filled with a simple, 
volatile fluid
such that the bulk, i.e., far away from the edge at $x=z=0$,
is occupied by the vapor phase with temperature $T$ and chemical potential
$\mu$. The values of $\mu$ are chosen such that the
bulk is close to liquid-vapor coexistence $\mu_{0}(T)$.
(Our considerations are equally applicable to the two fluid phases
of a phase separated binary liquid mixture.) Accordingly, under  the
influence of the substrate potential a liquidlike wetting film will
form at the planar surfaces leading to a meniscus near the edge. We
analyze the height of this meniscus and its shape as a function of
$\mu$ and $T$. This analysis is based on thermodynamic considerations
(Sec.II) and on a phenomenological interface model for the local
height of the meniscus (Sec.III). For a special choice of the effective 
interfacial potential we are able to calculate analytically the shape of the 
interface in the wedge explicitly (Sec.IV). By considering a
suitably chosen Landau-Ginzburg-Wilson theory for an order parameter 
in this geometry (Sec.V) we can discuss the range of validity for the
effective interface model studied in Sec.III. Our results
are summarized in Sec.VI.\\

\centerline{\bf {II. Macroscopic description}}
\renewcommand{\theequation}{2.\arabic{equation}} 
\setcounter{equation}{0}

In Secs.III and IV  we shall analyze microscopically 
the formation of a meniscus of a {\it volatile} liquid in thermal 
equilibrium with its vapor phase. Nonethless it will turn out
to be instructive to consider first in the following subsections
the corresponding macroscopic description of the problem in 
constrained equilibrium and to refer to the results obtained within
the standard theory of capillarity. \\

\centerline{\bf {A. Constrained equilibrium and thermodynamics}}

We consider a symmetric wedge (i.e., both walls consist of the same type 
of substrate material) with opening 
angle $2\varphi < \pi$ (see Fig.1(a)). It is cut off 
at a macroscopic height $H_{0}$. The lower part of the wedge is
filled with liquid; the corresponding wall-liquid, vapor-liquid, and 
wall-vapor surface tensions are denoted as
$\sigma_{wl}$, $\sigma_{lg}$, and $\sigma_{wg}$, 
respectively. The vapor and liquid phases are taken to be at bulk
equilibrium so that they can exchange volume with no cost in 
{\it bulk} free-energy. However, in this subsection we consider a 
constrained equilibrium such that the volume $V$ of the liquid phase 
is prescribed. The liquid
meniscus is enclosed by the walls at $z=|x|\cot \varphi$ and the
liquid-vapor interface is described by $f(x)$ (see Fig.1(a)). Here 
and in the following we discuss a two-dimensional wedge. Some 
aspects of the full three-dimensional case will be discussed in Subsec.IIB. 
In the present macroscopic description all length scales are much larger 
than atomic
scales so that the free energy $F$ of the system is determined by the
aforementioned {\it surface} tensions only. For a given configuration  
of the interface $f(x)$ one has (see Fig.1(a)) 
\begin{eqnarray}
F[f]\,=\,\sigma_{lg}\,\int
\limits_{{\mbox {$-x_{1}$}}}^{{\mbox {$\,\,\,x_{1}$}}}dx 
\,\sqrt{1\,+\,f_x^2(x)}
\,\,\,+\,\,\,2\,\frac{x_0\,-\,x_1}{\sin\varphi}\,\sigma_{wg}
\,\,\,+\,\,\,2\,\frac{x_1}{\sin\varphi}\,\sigma_{wl}\,\,\,
\end{eqnarray}
with $f_{x}=\frac{\mbox {$df$}}{\mbox {$dx$}}$. 
In Eq.(2.1) the artificial contributions to 
the free energy generated by the cutoff 
at $z=H_0$ have been omitted. The liquid-gas interface intersects the 
walls at $x= \pm\,x_1$ with  $f(\pm x_1) = x_1 \cot \varphi$. 
The surface tensions determine the contact angle $\Theta$ via the Young 
equation
\begin{eqnarray}
\cos\Theta\,=\,\frac{\sigma_{wg}\,-\,\sigma_{wl}}
{\sigma_{lg}}\,\,\,.
\end{eqnarray}
The difference $\Delta F[f] = F[f] - 
{\mbox {$2\,x_0\,\sigma_{wg}$}}/{\mbox {$\sin\varphi$}}$ 
of the energies of the wedge with  a given amount of liquid and of the wedge 
filled by the gas phase only follows from Eq.(2.1) by subtracting 
the term 
${\mbox {$2\,x_0\,\sigma_{wg}$}}/{\mbox {$\sin\varphi$}}$.
Accordingly the constrained equilibrium profile ${\bar f}$ minimizes the 
functional
\begin{eqnarray}
\Delta\,F^{\star}[f]\,=\,\sigma_{lg}\,\int
\limits_{{\mbox {$-x_{1}$}}}^{{\mbox {$\,\,\,x_{1}$}}}dx\,\left\{\
\sqrt{1\,+\,f_x^2(x)}
\,\,\,-\frac{\cos\Theta}{\sin\varphi}\,\,\,+\,\,\,
\frac{\lambda}{\sigma_{lg}}\,\left(f(x)\,-\,|x|\,\cot\varphi\right)
\right\}\,\,\, 
\end{eqnarray}
with the boundary conditions $f(\pm x_1)=x_1\cot \varphi $ and 
where the Lagrange multiplier $\lambda$ implements the constraint 
of constant liquid volume 
\begin{eqnarray}
V\,=\,\int\limits_{{\mbox {$-x_{1}$}}}^{{\mbox {$\,\,\,x_{1}$}}}dx 
\left(f(x)\,-\,|x|\,\cot\varphi\right)\,\,\,.
\end{eqnarray}
(In these formulae $f(x)$ is regarded as a single-valued function. 
In the case that $f(x)$ consists of two branches the corresponding separate 
analysis for each of the two branches leads to the same conclusions as the 
ones described below. Moreover it is necessary to assume that
$\Theta+\varphi<\pi$; otherwise neither the interface nor its branches can be
regarded as single valued functions.)  
According to the standard calculus of variations [46] 
the equilibrium
profile fulfills the differential equation
\begin{eqnarray}
\frac{{\bar f}_{xx}}{(1\,+\,{\bar f}_x^2)^{\frac{3}{2}}}\,=
\,\frac{\lambda}{\sigma_{lg}}\,\,\,
\end{eqnarray}
with the boundary conditions
\begin{eqnarray}
\frac{1\,\pm\,{\bar f}_x(\pm\,x_1)\,\cot\varphi}
{\left(1\,+\,{\bar f}_x^2(\pm\,x_1)\right)^{\frac{1}{2}}}
\,=\,\frac{\cos\Theta}{\sin\varphi}
\,\,\,.
\end{eqnarray}
Equation (2.6) is equivalent to the statement that the angle between
the liquid-gas interface and the wall (measured on the liquid
side of the interface) is equal to the contact angle $\Theta$
in the planar case. Thus the equilibrium profile is given by a
part of a circle whose radius 
$R=\frac{{\mbox {$\sigma_{lg}$}}}{{\mbox {$\lambda$}}}$
follows from the constant volume constraint (Eq.(2.4)) 
and intersects the walls at the contact angle $\Theta$.
For $\Theta<\frac{1}{2}\pi-\varphi$, the liquid-gas 
interface is concave, while for $\Theta>\frac{1}{2}\pi-\varphi$
the interface is convex; $\Theta=\frac{1}{2}\pi-\varphi$ 
corresponds to a flat interface. The center of the circle is located at 
$(x_c,z_c)$ with
\begin{eqnarray}
x_{c}=0\,\,,\,\,z_c\,=\,-\,R\,{\mbox {$\rm sign$}}\left(\Theta\,+\,\varphi\,-\,
\frac{1}{2}\,\pi\right)\,\frac{\cos\Theta}{\sin\varphi}\,\,\,,
\end{eqnarray}
where the radius $R$ is given by
\begin{eqnarray}
R\,=\,\sqrt{V}\left(\Theta\,+\,\varphi\,-\,\frac{1}{2}\pi\,
+\,\frac{\cos\Theta\,\cos(\Theta\,+\,\varphi)}{\sin\varphi}\right)
^{-\frac{1}{2}}\,\,\,.
\end{eqnarray}
The value of $z_c$ is always positive for a concave interface 
while for the convex one it can be either positive or negative, depending
on the value of the contact angle $\Theta$.
The intersection of the circle with the wall occurs at
\begin{eqnarray}
x_1\,=\,-R\,{\mbox {$\rm sign$}}\left(\Theta\,+\,\varphi\,-\,
\frac{1}{2}\,\pi\right)\,\cos(\Theta\,+\,\varphi)\,\,\,.
\end{eqnarray}
In the case that the liquid forms  a bridge connecting the walls the 
free energy is given by 
\begin{eqnarray}
F[f_1\,,\,f_2]\,=\,\sigma_{lg}\,\int
\limits_{{\mbox {$-x_{1}$}}}^{{\mbox {$\,\,\,x_{1}$}}}dx 
\,\sqrt{1\,+\,f_{1x}^2(x)}\,\,\,+\,\,\,\sigma_{lg}\,\int
\limits_{{\mbox {$-x_{2}$}}}^{{\mbox {$\,\,\,x_{2}$}}}dx 
\,\sqrt{1\,+\,f_{2x}^2(x)}\,\,\,+\,\,\,\nonumber\\
+\,\,\,2\,\frac{x_0\,-\,x_1\,+\,x_2}{\sin\varphi}\,\sigma_{wg}
\,\,\,+\,\,\,2\,\frac{x_1\,-\,x_2}{\sin\varphi}\,\sigma_{wl}\,\,\,
\end{eqnarray}
where $f_1(x)$ and $f_2(x)$ correspond to the 
upper and lower liquid-gas interface, respectively, 
while $\pm\,x_1$ and $\pm\,x_2$ denote the intersections with the walls. 
The minimization of the constrained free energy with
respect to $f_1$ and $f_2$ leads to the conclusion that
both interfaces are parts of the same circle whose center
is given by Eq.(2.7). For each of them the corresponding contact
angle (measured on the liquid side of the interface) has
the same value $\Theta$ as given by the contact angle for a planar
substrate. 

The bridge can exist if and only if $z_c$ is positive and larger  
than the radius of the circle, i.e., if  
$\Theta >\frac{1}{2}\pi+\varphi$. 
In this case the radius of the bridge has the value
\begin{eqnarray}
R\,=\,\sqrt{V}\,\left(2\,\Theta\,-\,\pi\,-\,\sin2\,\Theta\right)^{-\frac{1}{2}}
\,\,\,\,.
\end{eqnarray}
Thus, depending on the opening angle $\varphi$ and the contact angle $\Theta$, 
the possible shapes of a liquid meniscus can be summarized as follows (see 
Fig.2): 
\begin{eqnarray*}
\begin{array}{rl}
\Theta\,>\,\,\frac{1}{2}\,\pi\,+\varphi\,\,\, & {\mbox {: bridge}}\\
\,\,\,\,\,\,\frac{1}{2}\,\pi\,+\varphi\,\geq\,\Theta\,>\,\,
\frac{1}{2}\,\pi\,-\varphi\,\,\, & {\mbox {: single convex interface}}\\
\Theta\,=\,\,\frac{1}{2}\,\pi\,-\varphi\,\,\,
 & {\mbox {: flat interface}}\\
\frac{1}{2}\,\pi\,-\varphi\,>\,\Theta\;\;\;\;\;\;\;\;\;\;\;\;\;\;\;\;\;\;\;\;
 & {\mbox {: concave interface}}\,\,.
\end{array}
\end{eqnarray*} 
For these configurations the difference in free energy between the wedge 
being filled with a liquid of volume $V$ and vapor otherwise and the wedge 
being filled with vapor only is given by 
\begin{eqnarray}
\Delta\,F = {\mbox {sign}}\left(\Theta + \varphi - \frac{\pi}{2}\right)
2\sigma_{lg}\sqrt{V}\left(\Theta + \varphi - \frac{\pi}{2}
 + \frac{\cos\Theta \cos(\Theta + \varphi)}{\sin\varphi}\right)
^{\frac{1}{2}}\,,\\
{\mbox {single interface}}\,\,\,, \nonumber
\end{eqnarray}
and 
\begin{eqnarray}
\Delta\,F\,=\,4\,\sigma_{lg}\,\sqrt{V}\,\left(2\,\varphi\,+\,
\cos^2\Theta\,\cot\,\varphi\right)\,\left(2\,\Theta\,-\pi\,-\,
\sin\,2\,\Theta\right)^{-\frac{1}{2}}\,, \,{\mbox {bridge}}.
\end{eqnarray}
Any configuration with more than one bridge is disfavored by a higher 
free energy. $\Theta =\frac{1}{2}\pi+\varphi$ corresponds to the 
border case between a bridge and a convex interface for  which the lowest 
point of the bridge coincides with the corner of the wedge.  

From Eq.(2.12) one can infer that for 
$\Theta>\frac{1}{2}\pi-\varphi$, i.e., for a single convex interface 
or a bridge, $\Delta F$ decreases upon  decreasing the liquid volume down 
to the limiting value $V=0$.
On the other hand, for $\Theta<\frac{1}{2}\pi-\varphi$ the
free energy decreases upon increasing $V$, i.e., upon filling the whole 
wedge with liquid. Therefore 
\begin{eqnarray}
\Theta\,=\,\frac{1}{2}\pi\,-\,\varphi\,\,\,\,\,{\mbox {(wetting)}}
\end{eqnarray}
marks a filling transition for the wedge. This case corresponds to a flat 
interface (see Fig.2) and the corresponding free energy is independent of the 
volume of the liquid. At fixed temperature, i.e., $\Theta$ fixed, a wide 
wedge is covered  only by a microscopically thin liquid film which turns 
into a macroscopic meniscus upon narrowing the wedge beyond the critical
opening angle $\varphi_c=\frac{1}{2}\pi-\Theta$.
(These considerations can be extended to the case of a 
nonsymmetric wedge. The corresponding results are presented in Appendix A.) 
If the planar walls undergo a wetting transition the contact angle $\Theta(T)$ 
decreases as a function of temperature and vanishes at $T=T_{w}$, i.e., 
$\Theta(T=T_{w})=0$. 

Thus for a fixed opening angle Eq.(2.14) defines
implicitly a filling transition temperature $T_{\varphi}$ such that  
$\Theta(T_{\varphi})=\frac{1}{2}\pi-\varphi$. One has  
$T_{\varphi}<T_{w}$ and $T_{\varphi \rightarrow \frac{1}{2}\pi}=T_{w}$. 
This means that the wedge undergoes a filling transition at a temperature at 
which the outer parts of the wedge far away from the 
center still remain nonwet. The same conclusion has been reached
previously by Hauge from phenomenological considerations 
based on the Laplace and Young equations [34]. Here we obtain this
conclusion by analyzing the free energy of the system. 

However, as long as the volume of the liquid drop has a fixed value 
there is no phase transition and the free energy is an analytical
function of temperature and of the wedge opening angle. Close to  
$\varphi=\frac{1}{2}\pi-\Theta$ it can be expanded into a series 
of powers of $\left(\varphi-\left(\frac{1}{2}\pi-\Theta\right)\right)$. 
The dominant term of this expansion is 
\begin{eqnarray}
\Delta\,F\,=\,\simeq\,-\,2\,\sigma_{lg}\,\sqrt{V\,\cot\varphi}
\left(\varphi\,-\,\left(\frac{1}{2}\pi\,-\,\Theta\right)\right)\,\,\,. 
\end{eqnarray}
The aforementioned qualitative change of the interfacial 
shape and the change of the behavior of the free energy
upon increasing $V$
are traces of the filling transition which comes into play when
the fixed-volume constraint is removed and the system is at liquid-gas bulk 
coexistence of a grand canonical ensemble. The 
minimization of the functional $\Delta\,F[f]$ instead of
the functional $\Delta\,F^{\star}[f]$ leads to the following equation
for the equilibrium interfacial shape: 
\begin{eqnarray}
\frac{{\bar f}_{xx}}{(1\,+\,{\bar f}_x^2)^{\frac{3}{2}}}\,=0\,\,\,.
\end{eqnarray}
Thus the interface
is flat and horizontal. (In the general case of a nonsymmetric wedge
the interface still remains flat but not necessarily horizontal.)  
In the case of a nonconstrained system the surface free-energy is given by
\begin{eqnarray}
F\,(H)\,=\,\frac{2}{\cos \varphi}
\left[H_0\sigma_{wg}(T)\,+\,H(\cos\Theta(T_{\varphi})
\,-\,\cos\Theta(T))\sigma_{lg}(T)\right]\,\,\,,
\end{eqnarray}
where $H$ is the height of the flat and horizontal interface measured 
from the edge of the wedge. For $T<T_{\varphi}$ the above surface 
contribution is minimized by $H=0$, while for $T>T_{\varphi}$  
this contribution is minimized by the largest possible value for $H$. 
The latter case corresponds to a wedge completely filled by the
liquid. The relevant surface free energy has the following form: 
\begin{eqnarray}
F\,=\,\frac{2\,H_0}{\cos\varphi}\left\{
\begin{array}{ccl}
\sigma_{wg}(T)&\,\,\,\,\,\,, T\,\leq\,T_{\varphi},\,\,\,\,\\
 & \\
\sigma_{wg}(T)\,+\,
\left(\cos\Theta(T_{\varphi})\,-\,\cos\Theta(T)\right)
\sigma_{lg}(T)
&\,\,\,\,\,\,, T\,\geq\,T_{\varphi},\,\,\,.
\end{array}
\right. 
\end{eqnarray}
According to Eq.(2.18) $F$ is continuous at T\,=\,$T_{\varphi}$ 
but exhibits a break in slope as function of temperature. Thus the filling 
transition in a wedge is $-$ similar to a wetting transition on a planar 
substrate $-$ associated with a singularity in the corresponding {\it surface}
contribution to the free energy. But the relevant structural properties 
such as, e.g., the shape of the microscopically thin liquid film covering 
the wedge are determined by the {\it line} contribution to the free energy 
which is singular at  $T_{\varphi}$, too. This differs from the wetting 
transition for which all relevant structural properties are determined by the 
surface free energy alone. 

An analogous analysis excludes a bridge as a stable
configuration for the unconstrained system. Again 
both interfaces for such a bridge must be flat. 
If their heights are denoted as $H_1$ and $H_2<H_{1}$,
respectively, the free energy is given by
\begin{eqnarray}
F(H_1\,,\,H_2)\,=\,\frac{2}{\cos\varphi}\left[
H_0\,\sigma_{wg}(T)\,+\,\,\,
H_1\,\left(\cos\Theta(T_{\varphi})\,-\,\cos\Theta(T)\right)
\sigma_{lg}(T)\right.\,\,\,+\,\,\,\\
\left.H_2\,\left(\cos\Theta(T_{\varphi})\,+\,\cos\Theta(T)\right)
\sigma_{lg}(T)\right]\,\,\,.
\nonumber
\end{eqnarray}
Below $T_{\varphi}$ this expression is minimized by $H_{1}=0$ and 
$H_{2}=0$.  Above $T_{\varphi}$ the minimum corresponds to $H_{2}=0$  
while $H_1$ takes its maximal possible value.

In the above considerations we have discussed a wetting situation,
i.e., the bulk is occupied by vapor and the wedgelike
substrate prefers the
liquid. Our results can be easily mapped onto the corresponding
drying situation in which the bulk is occupied by liquid and
the wedgelike substrate prefers the vapor phase. In this case the
filling transition of the wedge occurs for
\begin{eqnarray}
\Theta\,=\,\frac{1}{2}\,\pi\,+\,\varphi\,\,\,\, {\mbox {(drying)}}\,. 
\end{eqnarray}
This is in accordance with Eq.(2.14) because liquid and gas are 
interchanged but the contact angle is in both cases taken to be the 
one of the 
liquid phase, i.e., in Eq.(2.14) $\Theta$ must be replaced by
$\pi\,-\,\Theta$. If the planar substrate supports the drying
transition, i.e., $\Theta(T\rightarrow T_d) = \pi$, the
wedge will be completely filled by the vapor phase for $T>T_{\varphi}$
where now $\Theta(T=T_{\varphi})=\frac{1}{2}\pi+\varphi$.
Again one has $T_{\varphi}<T_d$ and 
$T_{\varphi\rightarrow \frac{1}{2}\pi}=T_d$.  
Equation (2.20) states that, under drying conditions,
for all temperatures corresponding to $\Theta(T)<\frac{1}{2}\pi$
there is no opening angle $\varphi_c$ such that the wedge is
filled with vapor for $\varphi\,<\,\varphi_c$. On the other hand,
if $\Theta(T)>\frac{1}{2}\pi$ there is always a sufficiently
small opening angle $\varphi_c(T)=\Theta(T)-\frac{1}{2}\pi$
such that the wedge is filled up with vapor for $\varphi <
\varphi_c$.

This observation may be of relevance for preparing super 
water-repellent surfaces [47]. These are surfaces which exhibit a contact
angle $\pi\,>\,\Theta\,>\,\frac{1}{2}\,\pi$ if they are very smooth.
On the other hand, if the surfaces $-$ made of the same material $-$ are 
prepared such that they exhibit a very porous structure one
observes an apparent contact angle close to $\pi$ [47]. In the present
context this observation is in accordance with a filling of
(wedgelike) pores with vapor at temperatures at which the smooth
planar surface is not yet dry, i.e., $\Theta\,<\,\pi$.

Similar considerations hold for wetting conditions, i.e., when the
wedge is exposed to a vapor phase.
Equation (2.14) states that for all temperatures corresponding to $\Theta(T)
\,>\,\frac{1}{2}\,\pi$ there is no opening angle $\varphi_c$ such
that the wedge is filled with liquid for $\varphi\,<\,\varphi_c$.
On the other hand, if $\Theta(T)\,<\,\frac{1}{2}\,\pi$ there is
always a sufficiently small opening angle $\varphi_c\,(T)\,=\,
\frac{1}{2}\,\pi\,-\,\Theta(T)$ such that the wedge is filled with
liquid for $\varphi\,<\,\varphi_c$.\\

\centerline{\bf {B. Theory of capillarity}}

In the previous two subsections we have considered the constrained
equilibrium of fluid configurations which are translationally invariant 
along the $y$ direction of the edge of the wedge. Effectively this 
corresponds to a two-dimensional system. 
In a three-dimensional system, a fixed finite volume $V$ of the liquid 
must have a finite extension in the $y$ direction and thus the shape 
$f(x\,,\,y)$ of the liquid-vapor interface depends also on $y$. Based 
on the classical Young-Laplace-Gauss capillarity theory the equilibrium 
liquid-vapor interface 
configurations are surfaces of constant mean curvature meeting the
bounding walls with contact angle $\Theta$.

There exists a sizable body of literature devoted to the solution
of this problem for the present wedge geometry (see, e.g., Refs.[48-53] 
and references therein). The most recent account of the present
knowledge based on mathematically rigorous results and numerical
evidence is summarized in Fig.6 in Ref.[53]. 
 
For $\Theta\,>\,\frac{1}{2}\,\pi$ there are two possible liquid
configurations depending on the opening angle $2\,\varphi$.
For $\varphi\,\geq\,\Theta\,-\frac{1}{2}\,\pi\,>\,0$ an edge blob
forms; this is a part of a sphere such that the liquid is in contact
with both walls and the edge of the wedge. The shape of the liquid-vapor 
interface is convex. If the opening angle of the wedge is
reduced such that $\varphi\,<\,\Theta\,-\,\frac{1}{2}\,\pi\,>\,0$
the liquid looses contact with the edge of the wedge and a spherical
bridge connecting the two walls is formed. The transition between
these two distinct configurations occurs at $\varphi\,=\,\Theta
\,-\,\frac{1}{2}\,\pi\,>\,0$. This means that a preference of the planar
substrate for the vapor phase, i.e., $\Theta>\frac{1}{2}\pi$
implies a prefilling of the wedge with vapor for sufficiently
small opening angles of the wedge. 

For $\Theta\,<\,\frac{1}{2}\,\pi$ a tubular bridge between the
walls is not possible. For a wide wedge with $\varphi\,>\,\frac{1}{2}
\,\pi\,-\,\Theta\,>\,0$ one finds an edge blob in contact with both walls and
with the edge of the wedge. Upon decreasing the opening angle of the
wedge at $\varphi\,=\,\frac{1}{2}\,\pi\,-\,\Theta\,>\,0$ a transition
to edge spreading occurs which persists for $\varphi\,\leq\,\frac{1}{2}\,
\pi\,-\,\Theta\,>\,0$. Edge spreading by the liquid is not possible for 
$\Theta\,>\,\frac{1}{2}\,\pi$. This means that a preference of the planar 
substrate for the vapor phase, i.e., $\Theta \,>\,\frac{1}{2}\pi$ implies 
a prefilling of the wedge with vapor for sufficiently small opening angles 
of the wedge. These findings are in full accordance with the
free energy analysis of the effectively two-dimensional system discussed 
in Subsec.IIA. \\

\centerline{\bf{III. Description of filling transitions by an interface model}}
\renewcommand{\theequation}{3.\arabic{equation}} 
\setcounter{equation}{0}

Within the macroscopic description in Sec.II the type of substrate forming the 
wedge enters only summarily via the surface tensions $\sigma_{wg}$ and 
$\sigma_{wl}$. In the actual microscopic picture the fluid particles are exposed 
to the external substrate potential $V(x,z)$ exerted by the particles forming 
the 
wedge. The resulting full number density distribution $\rho(x,z)$ of the fluid 
particles can be determined, e.g., either by simulations or by density 
functional 
theory. However, in view of the considerable numerical challenges by such an 
approach so far only hard sphere fluids confined by hard walls have been studied 
in such full detail [40,41]. The accessible system sizes of the wedges which can 
be studied within these approaches are also severly limited. Moreover, without 
attractive interactions the hard-body systems do not exhibit filling 
transitions. \\

\centerline{\bf {A. Effective interface Hamiltonian}}

The study of effective models for the liquid-vapor interface exposed to an 
effective interface  potential [54,55], which takes into account the competition 
between the substrate potential and the interaction potential between the fluid 
particles, serves as a reasonable compromise between a full-fledged microscopic 
theory and the purely macroscopic picture. Within this effective approach the 
morphology of the liquid-vapor interface is determined by the effective 
interface 
Hamiltonian (see Fig.1(b)) 
\begin{eqnarray}
{\cal H}[l]=\int\limits_{\mbox {$-\infty$}}
^{\mbox {$\infty$}}dx\left\{\sigma_{lg}
\left(\sqrt{1+\left(\frac{df(x)}{dx}\right)^2}-
\sqrt{1+\cot^2\varphi}\right)+\frac{V(l(x) \sin\varphi)
\,-\,V(l_{\infty}\,\sin\varphi)}
{\sin\varphi}\right\}.
\end{eqnarray}
Since $z=|x|\cot\varphi$ is the position of the surfaces of the wedge, the 
local thickness of the liquid film measured vertically equals 
$l(x)=f(x)-|x|\cot \varphi$ and thus exhibits a cusp at $x=0$ with 
$l'(x \rightarrow \pm 0)=\mp \cot \varphi $. Thus the first term in Eq.(3.1) 
corresponds to the cost in free energy due to the increase of  the surface area 
of the liquid-vapor interface relative to its flat configuration at $|x|=\infty$ 
where $\left|\frac{\mbox {$df$}}{\mbox {$dx$}}(x \rightarrow \infty)\right|= 
\cot \varphi $ so that 
$ \sqrt{1+\cot^{2}(\varphi)}=\frac{\mbox {$1$}}{\mbox {$\sin\varphi$}}$. 
The second part of the 
Hamiltonian takes into account the effective interaction $V$ between the 
liquid-vapor interface and the wedge-shaped substrate relative to the 
configuration $l(|x| \rightarrow \infty)=l_{\infty}$. $V$ denotes the 
effective interface potential which we take to be of the same form as for a 
horizontal liquid-vapor interface interacting with the corresponding planar 
substrate [54,55]:
\begin{eqnarray}
V(L)\,=\,\sigma_{wl}\,+\,\sigma_{lg}\,+\omega(L)\,+\Delta\mu\,\Delta\rho\,L 
\end{eqnarray}
with $\omega(L\rightarrow\infty)=0$, $\Delta\mu=\mu_{0}(T)-\mu$, and 
$\Delta\rho=\rho_{l}-\rho_{g}$ where $\rho_{l}$ and  $\rho_{g}$ are the 
number density of the liquid and vapor, respectively. In Eq.(3.1) $V(L)$ is 
evaluated in a local approximation such that $L=l(x)\sin \varphi $ is the local 
thickness of the wetting film normal to the {\it near} surface of the substrate; 
therefore these potential terms are integrated with respect to 
$\frac{\mbox {$dx$}}{\mbox {$\sin \varphi$}}$, i.e., along the substrate surface 
(see Fig.1(b)). Thus 
Eq.(3.1) neglects the additional effective interaction of the liquid-vapor 
interface with the {\it distant} substrate surface. This approximation is 
expected to be valid for a rather open wedge (see below). The equilibrium film 
thickness on a planar substrate minimizes $V(L)$ and yields the actual 
substrate-vapor surface tension $\sigma_{wg}=V(l_{\infty}\sin \varphi) $ of the 
planar substrate. Due to the subtraction of those terms which correspond to the 
asymptotic behavior for $x\rightarrow \pm\infty$ the
Hamiltonian ${\cal H}[l]$ is finite for all 
configurations compatible with the boundary 
conditions. Therefore ${\cal H}[l]$ 
describes the {\it line contribution} to the free energy associated with the 
linear extension $L_{y}$ of the wedge in the $y$ direction. In the present 
mean-field theory we neglect the fluctuations of the interface along the $y$ 
direction so that $f$ depends on $x$ only and the line contribution carries 
simply a factor $L_{y}$ which has been omitted in Eq.(3.1). We emphasize that 
this consideration of the line contribution to the free energy represents the 
most important improvement over the macroscopic description which considers 
only surface contributions. Within mean-field theory the equilibrium shape 
${\bar f}(x)$ of the meniscus minimizes this line contribution: 
\begin{eqnarray}
\sigma_{lg}\,\frac{\frac{\mbox
{$d^{2}{\bar f}(x)$}}{\mbox
{$dx^{2}$}}}{\left(1\,
+\,\left(\frac{\mbox {$d{\bar f}(x)$}}
{\mbox {$dx$}}\right)^{2}\right)^{\frac{3}{2}}}\,=\,
V'(({\bar f}(x)-|x|\cot\varphi) \sin\varphi) \,\,\,,
\end{eqnarray}
\begin{eqnarray}
{\bar f}'(0)=0\,\,,\,\, {\bar f}(x\rightarrow \pm \infty)= 
l_{\infty}+|x|\cot \varphi \,\,. \nonumber\
\end{eqnarray}
Equation (3.3) is known as the so-called augmented Young equation and has been 
studied by Kagan et al. [39] but not with the view of filling phase transitions; 
recently these authors extended their analysis to study eye-shaped capillaries 
[56]. Due to the symmetry of the system we confine our subsequent analysis to 
$x \geq 0$. In terms of $l(x)$ the boundary conditions are 
$l(x \rightarrow \infty)=l_{\infty}\sin \varphi $ and $l'(x=0+)=-\cot 
\varphi$. 
Integrating Eq.(3.3) yields 
\begin{eqnarray}
\sigma_{lg}\, \left( 1\,-\,\frac{\sin \varphi + \cos \varphi 
\,\frac{\mbox {$d {\bar f}(x)$}}
{\mbox {$dx$}}} {\sqrt{1+\left(
\frac{\mbox {$d {\bar f}$}}{\mbox {$dx$}}\right)^{2}}}\right)\,= \,
V(\bar l(x)\,\sin \varphi) \,-\,V(l_{\infty}\,\sin \varphi)\,\,.
\end{eqnarray}
With ${\bar f}'(0)=0$ and $l_0
\equiv {\bar l}(x=0)$ one obtains from Eq.(3.4) 
\begin{eqnarray}
\frac{\Delta V}{\sigma_{lg}}\,\equiv \frac{1}{\sigma_{lg}}\,
\left[V(l_{0}\,\sin \varphi)-V(l_{\infty}\,\sin \varphi)\,\right]\,= 
1-\sin \varphi \,\equiv {\bar v}(\varphi)\,\,.
\end{eqnarray}
Equation (3.5) is an implicit algebraic equation for the filling height $l_0$ in 
the center of the wedge in terms of the opening angle $\varphi$ and the wetting 
properties of the planar system, i.e., $\sigma_{lg}$, $V(L)$, and thus 
$l_{\infty}$. 
At liquid-vapor coexistence $V(L\rightarrow \infty)=\sigma_{wl}+\sigma_{lg}$  
(Eq.(3.2)) with $\frac{\mbox {$\Delta V$}}
{\mbox {$\sigma_{lg}$}}(L \rightarrow \infty)=1-\cos 
\Theta $ (Eq.(2.2)) so that the condition for the 
filling transition ${\bar l}(0) \rightarrow \infty$ can be 
expressed in terms of the contact angle $\Theta$ of the planar system 
(Eq.(2.2)):
\begin{eqnarray}
\cos\,\Theta(T_{\varphi})\,=\, \sin \varphi \,\,\,\, {\mbox {or}} \,\,\,\, 
\Theta(T_{\varphi})\,
=\,\frac{1}{2}\pi\,-\,\varphi\,\,.
\end{eqnarray}
$T_{\varphi}$ is the lowest temperature for which $l_0=\infty$. Thus our 
microscopic approach confirms the results for $T_{\varphi}$ as predicted by the 
macroscopic theory in Sec.II. (This conclusion even holds if the integrand in 
Eq.(3.1) is supplemented by a term proportional to the mean curvature of the 
interface.) It is rather satisfactory to see {\it explicitly} that a microscopic 
theory for the line contribution to the free energy renders the same value for 
the 
filling transition temperature $T_{\varphi}$  as the macroscopic considerations 
based on the surface free energies. On the other hand this is to be expected 
because at $T_{\varphi}$ the surface free energy of the wedge is nonanalytic 
(see Eq.(2.18)) so that the line free energy has to follow suit. This is 
analogous 
to the fact that surface free energies are  nonanalytic at bulk transitions. 

Before we turn to a closer analysis of the filling transition we note that in 
the 
special case of a wide open wedge, i.e., $\varphi = \frac{1}{2}\pi- \epsilon$ 
with 
$\epsilon << 1$ the effective interface Hamiltonian in Eq.(3.1) reduces to (see 
Appendix B) 
\begin{eqnarray}
{\cal H}[f]\,=\,\int\limits_{\mbox {$-\infty$}}
^{\mbox {$\infty$}}\,dx\left\{\frac{\sigma_{lg}}{2}\, 
\sin \varphi \, 
\left[\left(\frac{df(x)}{dx}\right)^2\,-\,
{\cot}^{2}\varphi\right]\,+\,\right.\\ \nonumber \left. 
\frac{V(l(x)\,\sin\varphi)\,-\,
V(l_{\infty}\,\sin\varphi)}
{\sin\varphi}\right\}\,\,\,,\,\,\, 
\left|\frac{1}{2}\,\pi\,-\,\varphi\right| \ll 
1\,\,,
\end{eqnarray}
so that the equilibrium profile ${\bar f}(x) ={\bar l}(x)+|x|\cot \varphi$ is 
determined by 
\begin{eqnarray}
\sigma_{lg}\,\sin\varphi \,\frac{d^{2}{\bar l}(x)}{dx^2}\,=\,V'({\bar l}(x)
\sin \varphi)\,,\,\,x>0\,.
\end{eqnarray}
Upon integration one obtains with $\frac{\mbox {$d{\bar l}$} }{\mbox {$dx$}}
(x = +\infty)=0$ 
\begin{eqnarray}
\frac{1}{2}\,\sigma_{lg}\, {\sin}^2 \varphi\, 
\left(\frac{d{\bar l}(x)}{dx}\right)^2\,=\,
V({\bar l}(x) \sin \varphi)\,-\,V(l_{\infty} \sin \varphi)\,.
\end{eqnarray}
Thus for the filling height $l_0$ in the center of the wedge one finds due to 
$\frac{\mbox {$d{\bar l}$}}{\mbox {$dx$}}(x \rightarrow \pm 0)=\mp 
\cot \varphi$ 
(compare Eq.(3.5) 
\begin{eqnarray}
\frac{\Delta V}{\sigma_{lg}}\,=\,\frac{1}{2}\,\cos^2 \varphi \,\equiv 
v(\varphi)\,.
\end{eqnarray}
According to the relations $\,\frac{1}{2}\cos^2 \varphi =(1-\sin \varphi)\,
\cos^2(\frac{1}{2}(\frac{1}{2}\pi-\varphi))\,$ and $\,1- \sin \varphi=2{\sin}^2
(\frac{1}{2}(\frac{1}{2}\pi-\varphi))$ the approximate implicit equation 
(3.10) 
for $l_0$ differs from the corresponding full equation (3.5) only by terms of 
the 
order ${\epsilon}^4$: $v(\varphi)={\bar v}(\varphi)(1+O({\epsilon}^2))$ and 
${\bar v}(\varphi)=O({\epsilon}^2)$. This is also true for the equation  for the 
filling transition temperature $T_{\varphi}$ (compare Eq.(3.6)) 
\begin{eqnarray}
\cos\,\Theta(T_{\varphi})\,=1-\frac{1}{2} \cos^2 \varphi\,\,,\,\, 
\left|\frac{1}{2}\,\pi\,-\, \varphi \right| \ll 1 \,\,\;, \\ 
= \sin \varphi + 2\,\sin^{4}\left(\frac{1}{2}\left(
\frac{1}{2}\,\pi-\varphi\right)\right)\,. \nonumber 
\end{eqnarray}
Thus we conclude that Eqs.(3.7)-(3.11) 
are reliable approximations for a wide open wedge.
\newpage

\centerline{\bf {B. Filling height}}

In the implicit equation (3.5) and its approximation (Eq.(3.10)) the right hand 
side ${\bar v}(\varphi)$ and $v(\varphi)$, respectively, do not depend on $l_0$; 
furthermore the left hand side 
$\Delta V/\sigma_{lg}$ remains unchanged 
upon the open wedge approximation. This facilitates a transparent graphical 
solution for the filling height $l_0$ as shown in Figs.3 and 4. As anticipated 
the 
filling height $l_0$ is larger than the wetting film thickness $l_{\infty}$ on 
the 
walls of the wedge far away from the edge of the wedge. At gas-liquid 
coexistence, 
i.e., $\Delta \mu =0$ the filling height $l_0$ diverges smoothly for 
$T \nearrow T_{\varphi}$ in the case of critical wetting and jumps to a 
macroscopic value in the case of first-order wetting of the corresponding planar 
substrate. At $T = T_{\varphi}$ the wetting film thickness $l_{\infty}$ 
asymptotically far away from the center of the wedge remains finite. Thus within 
the present model we find that the wedge does indeed undergo a filling 
transition 
at  $T_{\varphi}$ and that the order of the filling transition agrees with the 
order of the wetting which takes place at $T_{w} > T_{\varphi}$ on the 
corresponding planar substrate  for which $\Theta(T \rightarrow T_{w})=0$. 

As will be discussed in more detail in the following subsection the filling 
height 
$l_0$ diverges along isotherms $\Delta \mu \rightarrow 0$ for $T > T_{\varphi}$. 
In that case in Eq.(3.5) the asymptotic behavior of $V(L \rightarrow \infty) = 
\Delta \mu \Delta \rho L + \sigma_{wl}+\sigma_{lg}$ can be used. Together 
with 
$V(l_{\infty}\sin\varphi)=\sigma_{wg}$ and Eq.(2.2) this leads to $\Delta V / 
\sigma_{lg} \rightarrow \Delta \mu \Delta \rho l_{0}  \sin \varphi /
\sigma_{lg}  +1 -\cos \Theta$ so that from Eq.(3.5) one obtains 
\begin{eqnarray}
l_{0}(\Delta\, \mu\, \rightarrow \,0\,\,,\,\, T\, >\, T_{\varphi})\,=\,
\frac{\cos \Theta\, - \,\sin \varphi}
{\sin \varphi}\, \frac{\sigma_{lg}}{\Delta \rho}\,\frac{1}{\Delta \mu}\,\,\,.
\end{eqnarray}
We note that the form of Eq.(3.12) is valid irrespective of the order of the 
filling transition and irrespective of the range of the molecular forces. The 
latter 
enter only indirectly via $\Theta$, $\sigma_{lg}$, and $\Delta \rho$. Equation 
(3.12) is in full agreement with the macroscopic description in Ref.[34] and the 
numerical results above $T_{w}$ in Ref.[33]. \\

\centerline{\bf {C. Line tension}}

As apparent from Figs.3 and 4 the implicit equation (Eqs.(3.5) and (3.10)) for 
the 
filling height $l_0$ yields two or even more solutions. For  $l_{0} < \infty$, 
i.e., for $T < T_{\varphi}$ with $T_{\varphi}$ defined by Eqs.(2.14) and (3.6) 
or 
for $\Delta \mu > 0$, the equilibrium solution is that one whose corresponding 
profile $l(x)$ with $l(0)=l_{0}$ minimizes the line contribution of the free 
energy 
(Eqs.(3.1) or (3.7)). For $l_{0} < \infty$ all competing configurations have the 
same surface free energy. At coexistence, i.e., for $\Delta \mu =0$, $l_{0}$ can 
become 
macroscopically large. In that case one has to consider both the line and the 
surface contribution such that if one solution has a lower surface energy it 
wins 
out irrespective of the behavior of the line tension; the line tensions matter 
only if the surface free energies are equal. (By construction the bulk free 
energies of all configurations are always the same.) Nonetheless, in any case it 
is 
interesting to study the thermodynamic behavior of line tension.  

From Eq.(2.18) one infers that at coexistence the temperature dependence of the 
surface free energy density is given by $\sigma_{wg} + \delta \sigma (T)$ with 
$\delta \sigma(T < T_{\varphi})=0$ and $\delta \sigma(T > T_{\varphi}) < 0$ due 
to $\sigma_{lg}(T) >0 )$. This implies that for $T > T_{\varphi}$ the filled 
wedge 
exhibits a surface free energy which is lower than the surface free energy of 
the 
unfilled wedge extrapolated to $T > T_{\varphi}$. This holds independently of \
whether $\sigma_{wg}(T)$ is an increasing or decreasing function of $T$. Thus 
we conclude that $l_{0}=\infty$ for all thermodynamic states ($\Delta \mu = 0, 
T > T_{\varphi}$). 

For all other thermodynamic states with $\Delta \mu \geq 0$  one has 
$l_{0} < \infty$ and the line contribution to free energy (Eq.(3.7)) of the 
corresponding profile $f(x)$ can be determined explicitly. From Eq.(3.9) one 
has \ 
$\frac{\mbox {$dl$}}{\mbox {$dx$}} 
= \mp (\sqrt{2}/(\sqrt{\sigma_{lg}} \sin \varphi))
\sqrt{V(l(x) \sin \varphi) - V(l_{\infty} 
\sin \varphi)}$ for $x > 0$ (upper sign) or $x < 0$ (lower sign). 
With 
$\frac{\mbox {$df$}}{\mbox {$dx$}}
= \frac{\mbox {$dl$}}{\mbox {$dx$}} \pm \cot \varphi$ for $x > 0$ (upper sign) 
or $x < 0$ (lower sign) 
the insertion into Eq.(3.7) yields for the line tension $\eta$ 
\begin{eqnarray}
\eta = 2 \sqrt{2} \sigma_{lg}\, A 
\end{eqnarray}
with 
\begin{eqnarray}
A=\int\limits
_{\mbox {$l_{\infty}$}}^{\mbox {$l_{0}$}} 
dl \left\{\sqrt{\frac{V(l \sin\varphi)-
V(l_{\infty} \sin\varphi)}{\sigma_{lg}}}- \sqrt{\frac{V(l_{0} \sin\varphi)-
V(l_{\infty} \sin\varphi)}{\sigma_{lg}}} \right\}
\end{eqnarray}
where $l_{0}$ is a solution of Eq.(3.10) and $l_{\infty}$ minimizes $V(L)$. The 
quantity $A$, which has the dimension of a length and which gives the line 
tension 
up to the positive prefactor $2 \sqrt{2} \sigma_{lg}$, is the area between the 
curves \ 
$\sqrt{(V(l \sin\varphi)-V(l_{\infty} \sin\varphi))/\sigma_{lg}}$, 
$\sqrt{v(\varphi)}=\sqrt{(V(l_0 \sin\varphi)-
V(l_{\infty} \sin\varphi))/\sigma_{lg}}$ (see Eq.(3.10)), and $l=l_{\infty}$ 
(see Figs.5 and 6). Note that obviously the curves $\Delta V$ and $v(\varphi)$ 
intersect at the same position $l_0$ as $\sqrt{\Delta V}$ and 
$\sqrt{v(\varphi)}$. 
From this graphical interpretation one infers immediately that only the 
solutions 
$l_{0} > l_{\infty}$ have to be taken into account; as expected the filling 
height 
in the center of the wedge is larger than the thickness of the wetting film far 
outside and these solutions are the ones which increase for 
$T \nearrow T_{\varphi}$.

One can infer the filled state for $\Delta \mu =0$ and $T > T_{\varphi}$ by 
considering isotherms $\Delta \mu \rightarrow 0$ for $T > T_{\varphi}$ 
(see Eq.(3.12)). Figure 7(a) describes the continuous filling of the wedge for 
$T > T_{\varphi}$ and $\Delta \mu \rightarrow 0$ in the case of a critical 
wetting transition. For an underlying first-order wetting transition Fig.7(b) 
demonstrates that, along an isotherm with $T > T_{\varphi}$, at 
${\Delta \mu}_{pf}(T)$ one encounters a thin-thick transition of the filling 
height which is {\it not} accompanied by a phase transition in the structure 
of the wetting film far away from the center of the wedge. We call this phase 
transition a {\it prefilling transition}. Once this prefilling transition locus 
has 
been passed, the filling height $l_0$ diverges continuously for 
$\Delta \mu \rightarrow 0$ (see Eq.(3.12)). This confirms, within the present 
interface model, the expectation that the thermodynamic states 
$(\Delta \mu =0, 
T > T_{\varphi})$ correspond to a filled wedge. The order of the filling 
transition at coexistence is linked to the order of the underlying wetting 
transition.  

From Fig.7(b) one infers that a jump in the thickness $l_{\infty}$ of the 
wetting film far away from the center of the wedge upon crossing the prewetting 
line enforces a discontinuity in the line contribution to free energy 
associated with a discontinuity of the whole profile $l(x)$, however such, 
that $-$ surprisingly $-$ $l_{0}=l(x=0)$ happens to change only smoothly. On 
the other hand, upon crossing the prefilling line $l_0$ changes discontinuously 
without a change in $l_{\infty}$. This behavior 
confirms the general picture that a nonanalyticity at one thermodynamic level 
(bulk, 
surface, line, ...) induces nonanalyticities at the same locus at all 
subdominant 
thermodynamic levels and that each thermodynamic level can develop new 
nonanalyticities at loci where all higher thermodynamic levels are strictly 
analytic: 
the phase boundaries in the bulk free energy are lines of 
nonanalyticities both for 
the surface and the line contributions, and the prewetting line nonanalyticity 
of the 
surface free energy is also the locus of  nonanalyticities in the line tension.
On 
the other hand the bulk free energy is analytic along the prewetting line and 
the 
bulk and the surface free energy are analytic along the prefilling line. 

If the  wedge fills, on each side of the wedge a three-phase contact line 
between 
the substrate, liquid, and vapor is forming. In the limit 
$l_{0}\rightarrow \infty$ these two contact lines become independent and each of 
them reduces to the structure of a single three-phase contact line on a planar 
substrate. Therefore  one expects that in the limit $l_{0}\rightarrow \infty$ 
the line tension given by Eqs.(3.13) and (3.14) should reduce to twice the line 
tension ${\eta}_{plan}$ of the corresponding single contact line on a planar 
substrate:
\begin{eqnarray}
\eta(l_{0} \rightarrow \infty)\,=\, 2 \sqrt{2 \sigma_{lg}}\,\int
\limits_{l_{\infty}}^{\infty}dl\,
\left\{\sqrt{V(l \sin\varphi)-V(l_{\infty} \sin\varphi)}\,-\,\sqrt{
\frac{\sigma_{lg}}{2}}\cos \varphi\right\} \\ \nonumber =\, 
2\,{\eta}_{plan}\,+\,
O\left(\left(\frac{1}{2}\,\pi-\varphi\right)^2\right)
\end{eqnarray}
where within the present interface model ${\eta}_{plan}$ 
is indeed given by [57]
\begin{eqnarray}
{\eta}_{plan}\,=\,\sqrt{2 
\sigma_{lg}}\int\limits_{l_{\infty}}^{\infty}dl\,\left\{\sqrt{V(l)
-V(l_{\infty})}\,-\,\sqrt{-S}\right\}
\end{eqnarray}
with the spreading coefficient $S=\sigma_{wg}-\sigma_{lg}-\sigma_{wl}=-
\sigma_{lg}(1-\cos \Theta)$ equal to 
$-\frac{1}{2}\sigma_{lg}\,{\cos}^{2}\varphi$ 
for $T = T_{\varphi}$. However, within a full theory one expects that 
$\eta(l_{0}\rightarrow \infty)=2{\eta}_{plan}+{\eta}_{l}$ where ${\eta}_{l}$ 
is the line tension of a wedge filled with liquid, including the liquid phase as 
boundary condition in the bulk limit. But this latter contribution is not 
contained in the present {\it interface} model. \\

\centerline{\bf {D. Phase diagram}}

In order to translate the general features of the line tension discussed
in the previous subsection into an actual phase diagram one has to
specify the functional form of the effective potential, i.e., 
$\omega(L)$ (see Eq.(3.2)). In the spirit of the square gradient
expression (Eq.(3.7)), which is applicable for systems with 
short-ranged forces [58], we choose the generic form obtained by Fisher and 
Jin [59]  
\begin{eqnarray}
\omega(L)\,=\,W\,\exp\left(-\frac{L}{\xi}\right)+
U\,\left(1-C\,W^2\frac{L}{\xi}\right) 
\exp\left(- \frac{2L}{\xi}\right)+\Delta \mu
\Delta \rho L\,\,\,.
\end{eqnarray}
Since at present we are aiming for the generic, possible features of the
phase behavior in wedges we refrain from studying the specific
effects due to an algebraic decay of $\omega(L\rightarrow \infty)$
as it is characteristic for actual fluids governed by dispersion
forces [15]; this is left to future studies. In Eq.(3.17) $\xi$ is the
correlation length in the bulk of the wetting phase, i.e., the liquid
phase.

For $C\leq 0$ Eq.(3.17) yields a continuous wetting transition
at coexistence $\Delta \mu = 0$ if $W(T<T_w)<0$,
$W(T=T_w)=0$,$\,\,W(T>T_w)>0$, and $U(T_w)>0$.
In this case the phase diagram consists of the line of first-order
gas-liquid bulk transitions at $\mu=\mu_0(T)$ and of the 
temperatures $T_c$, $T_w$, and $T_{\varphi}$ as determined by Eq.(3.11)
on that line corresponding to the bulk, surface, and line 
nonanalyticities, respectively. For $\mu\neq\mu_0(T)$ there are no
nonanalyticities.

For $C>0$, $W(T)>0$, and $U>0$ Eq.(3.16) describes
a first-order wetting transition at $T=T_w$ due to a decrease of 
$W(T)$ for $T\nearrow T_{w}$ such that $W(T_w)=W_w(C,U) > 0$. For
reasons of simplicity we take $U$, $C$, and $\xi$ as constant and 
consider a linear temperature dependence of
$W(T)=W_w+W_0(T_w-T)/T_w$, $\,\,W_0>0$.
The chemical potential difference can be expressend in terms of 
the dimensionless variable $(\Delta\mu\Delta\rho
\xi)/W_w$.  Figure 8 illustrates the phase diagram 
predicted by this model. The thick lines indicate the bulk singularities
at $\mu\,=\,\mu_0$, which for simplicity we have taken to be
temperature independent, and the prewetting line emanating from
$T_w$ and ending at the prewetting critical point $C_{pw}$. 
The prewetting line $T_{pw}(\Delta\,\mu)$ (or $\Delta\mu_{pw}(T)$)
joins the gas-liquid coexistence curve tangentially [60]
such that $T_{pw}(\Delta\mu\rightarrow 0)-T_w
\sim\Delta\mu\ln\Delta\mu$ in accordance with the exponential
decay of $\omega(L\rightarrow \infty)$. On the present scale
of Figs.8 and 9 this tangential approach is not visible.
The first-order filling transition occurs at $T_{\varphi}
<T_w$ such that $T_{\varphi}$ approaches $T_w$ for $\varphi
\rightarrow \frac{1}{2}\pi$. 
This infinite jump from a microscopic 
filling height to a macroscopic height at coexistence is reduced
to a finite discontinuity off coexistence $\mu<\mu_0$
forming a prefilling line $\mu_{pf}(T)$ which joins the
gas-liquid coexistence curve also tangentially. From our numerical 
analysis we find $\mu_{pf}(T\searrow T_{\varphi})
-\mu_0\sim a(\Delta T)^2
+b(\Delta T)^4+...\,\,$ with $\Delta T=T-T_{\varphi}$.
The thin-thick jump of the filling heigth across the prefilling
line diverges  $\sim(\Delta\mu)^{-1}$ for $\mu\rightarrow\mu_0$
(see Eq.(3.12)) and vanishes upon approaching the prefilling
critical point $C_{pf}$.
For sufficiently narrow wedges the prefilling lines are completely
below the prewetting line and shorter than the latter. Upon
increasing the opening angle of the wedge the prefilling line
slides into the prewetting line thereby breaking up into two
pieces: one between $T_{\varphi}$ and a so-called extraordinary
point denoted as $E^{(1)}$ and another between a second extraordinary
point $E^{(2)}$ and the prefilling critical point $C_{pf}$.
In the limit $\varphi\rightarrow\frac{1}{2}\pi$ 
these two pieces 
shrink to zero, such that $E^{(1)}\rightarrow w$ and $E^{(2)}
\rightarrow C_{pw}$. Figure 9 summarizes the type of the phase
transitions in the meniscus shape of the liquid in the wedge across
the various phase boundaries for a very wide wedge.
Crossing the pieces of the prefilling line along the paths 1 and 2
leads to a discontinuous increase of the filling height of the wedge
but does not change the thickness of the wetting layer far away
from the center of the wedge. A subsequent crossing of 
the prewetting line leads to a discontinuous increase of the
thickness of the wetting film and a discontinuous
change of the meniscus profile $l(x)$ but such that $-$ surprisingly $-$
just the filling height $l_0=l(x=0)$ increases continuously
(paths 3 and 4). Along path 5 both the thickness
of the wetting film and the filling height change discontinuously;
that happens only upon crossing the prewetting line.
According to Fig.9 for suitable model systems along an
isotherm $\Delta\mu\rightarrow 0$ with a temperature
$T_{\varphi}<T<T_w$ one observes a reentrant prefilling
of the wedge. For large undersaturations the filling height
is very small. It increases until the prefilling line connecting
$C_{pf}$ and $E^{(2)}$ is reached. Crossing it leads to a
discontinuous decrease of the filling height upon decreasing
$\Delta\mu$. A further decrease of $\Delta\mu$ leads
again to an increase of the filling height which jumps to
an even larger value upon crossing the prefilling line
connecting the filling point $f$ and the extraordinary point
$E^{(1)}$. Once this prefilling line has been passed the filling
height diverges continuously for $\Delta\mu\rightarrow 0$
(see Eq.(3.12)).\\

\centerline{\bf {E. Fluctuations}}

Upon crossing the prefilling line the filling height $l_0$ jumps
from a value $l_0^{<}$ to a larger value $l_0^{>}$ so that a volume
$\Delta V$ proportional to $\frac{1}{2}((l_0^{>})^2-
(l_0^{<})^2)L_0\tan\varphi$ is transformed from gas
into liquid; $L_0$ is the linear extension of the wedge in $y$ 
direction. (The above expression corresponds to flat menisci at
height $l_0^{<}$ and $l_0^{>}$, respectively.)
Thus effectively $\Delta\rho\Delta V$ particles
participate in this phase transition. Therefore in the thermodynamic
limit $L_0\rightarrow\infty$ only at coexistence, i.e., 
for the filling transition with $l_0^{>}=\infty$, this phase
transition corresponds to a true two-dimensional system which
can indeed support a phase transition at $T>0$.
However, along the prefilling line $l_0^{>}$ is finite so that
in that case the system is quasi-one-dimensional and cannot undergo
a true phase transition for realistic interaction potentials.
Therefore we conclude that the prefilling lines, as predicted
by the mean-field theory described in the previous subsections,
are wiped out by fluctuations in space dimensions $d=3$.
Only the filling transitions {\it at} coexistence, which can be either
continuous or first-order, are true phase transitions. These
conclusions are in accordance with considering the critical 
points $C_{pw}$ of the prewetting line and $C_{pf}$ of the
prefilling line (see Figs. 8 and 9). Whereas $C_{pw}$ belongs to
the Ising universality class in $d=2$, $C_{pf}$ would belong
to the Ising universality class in $d=1$ and thus cannot exist.

One can identify the type of fluctuations which wipe out
the prefilling line. If it would exist, by imposing suitable
boundary conditions at the two ends of the groove for thermodynamic
states at the prefilling line one could generate a stable
interface perpendicular to the $y$ direction of the wedge which 
smoothly interpolates between a portion of the wedge filled up to
$l_0^{<}$ and another portion filled up to $l_0^{>}$.
However, in this quasi-one-dimensional system the fluctuations
of the filling height $l_0$ along the $y$ direction, which 
are not captured by mean-field theory, are so strong that for this
thermodynamic state at the presumed prefilling line the interface
configuration in the wedge breaks up into many domains with $l_0^{>}$
and $l_{0}^{<}$, respectively, whose positions fluctuate strongly.

However, close to liquid-vapor coexistence $l_0^{>}$ is very large
so that overturning a $l_0^{<}$-domain into a $l_0^{>}$-domain
and vice versa becomes increasingly improbable. Therefore at the 
prefilling line as function of $T$ or $\mu$ the filling
height $l_0(T,\mu)$ changes rapidly but smoothly between $l_0^{>}$
and $l_0^{<}$ such that along the prefilling line for
$\mu\rightarrow\mu_0$, i.e., $l_0^{>}\rightarrow\infty$
this crossover between $l_0^{>}$ and $l_0^{<}$ becomes steeper
and is confined to a vanishingly narrow region around the prefilling
line so that the true phase transition at the filling transition 
at coexistence is restored for $\Delta\mu\rightarrow 0$.

The width of this smooth transition region of $l_0(T,\mu)$
at the prefilling line can be estimated on the basis of the
finite-size scaling theory for first-order phase transitions [61].
These results can be adapted to the present problem following
 the line of arguments in Sec.4 in Ref.[62] where the corresponding
smearing out of the prewetting line on a cylindrical substrate of 
radius $r_0$ has been analyzed. Up to a pre-exponential factor, which depends 
inter alia on the details of the effective interface potential,
the temperature range $\delta\,T$ within which $l_0(T,\mu)$
crosses over smoothly from $l_0^{>}$ to $l_0^{<}$ is given by
\begin{eqnarray}
\frac{\delta\,T}{T_{\varphi}}\,\sim\,\exp\left(-\frac{\mbox
{${\bar \kappa}\,\Sigma$}}
{\mbox {$k_B\,T_{\varphi}$}}\right)\,\,\,,
\end{eqnarray}
where ${\bar \kappa}$ is a numerical factor of order unity and
$\Sigma$ is the energy required for the formation of a domain
wall between a $l_0^{>}$-domain and a $l_0^{<}$-domain. 
As a crude estimation we approximate $\Sigma$ by
$2\sigma_{lg}\left(l_0^{>}\right)^2\tan\varphi$. 
(Here the effective width $2l_0^{>}\tan\varphi$
of the wedge replaces the cylinder radius $r_0$ in
Ref.[62]; as in Ref.[62] the line tension $\Sigma_l$ introduced
there is appoximated by $l_0^{>}\sigma_{lg}$.) 
Since $\frac{\mbox {$\xi^2\sigma_{lg}$}}{\mbox {$k_BT_{\varphi}$}}$
is of order unity [62], where $\xi$ is the bulk correlation
length, we finally arive at the estimate
\begin{eqnarray}
\frac{\delta\,T}{T_{\varphi}}\,\sim\,
\exp\left(-\kappa\,\left(\frac{l_0^{>}}
{\xi}\right)^2\tan\varphi\right)\,\,\,.
\end{eqnarray}
As soon as $l_0^{>}$ becomes significantly larger than $\xi$
the temperature region $\delta\,T$ for the smooth crossover is
vanishingly small. Since $l_0^{>}$ diverges as $(\Delta\mu)^{-1}$
for $\Delta \mu\rightarrow 0$ along the prefilling
line we conclude that close to liquid-vapor coexistence
the difference between a true first-order thin-thick transition
for $l_0$ cannot be experimentally distinguished from the actual
smooth but very steep crossover. In this sense the prefilling
line as obtained by mean-field theory remains an experimentally
accessible line of (quasi-)nonanalyticities. Only close to
$C_{pf}$ this smearing out of the prefilling line becomes
effective. There, in Eq.(3.19) $l_0^{>}$ must be replaced by
$\l_0^{>}-l_0^{<}$. \\ 

\centerline{\bf IV. {Meniscus shape}}
\renewcommand{\theequation}{4.\arabic{equation}} 
\setcounter{equation}{0}

So far we have discussed the configuration in the wedge only in
terms of its key characteristic feature, i.e., the filling
height $l_0$ (see Subsec.IIIB). The more detailed information
about the full meniscus shape requires to solve Eq.(3.3) or its 
approximate version given by Eqs.(3.8) and (3.9). Whereas 
the former typically requires a numerical solution, the implicit
solution $l(x)$ of the latter reduces to an integration: 
\begin{eqnarray}
-\,\sqrt{\frac{1}{2}\,\sigma_{lg}}\,\sin\varphi\,
\int\limits_{\mbox {$l_0$}}^{\mbox {$l(x)$}}
\frac{d{\hat l}}{\sqrt{V({\hat l}\,\sin\varphi)\,-\,V
(l_{\infty}\sin\varphi)}}\,=\,x\,\,\,\,\,\,,\,\,\,\,\,\,
x\,>\,0\,\,\,,
\end{eqnarray}
where $l_0$ is the equilibrium solution of Eq.(3.10) and
$l(-x)=l(x)$. (Here and in the following we drop the overbar 
which indicates the minimum of Eq.(3.7).)
For a given effective interface potential $V(L)$ Eq.(4.1) can
readily be solved numerically.

For the model given in Eq.(3.17) it turns out that one can
obtain {\it explicit} solutions for $C=0$ and $\Delta \mu=0$
describing a {\it critical wetting} transition of the corresponding
planar substrate, i.e., $W=W_0\left(T_w-T\right)/
T_w$ where here $W_0<0$ and $U>0$. Within this 
model one has
\begin{eqnarray}
l_{\infty}\,=\,\frac{\xi}{\sin\varphi}\,\ln\left(\frac{2\,U}
{|W_0|\,t}\right)\,=
\frac{\xi}{\sin\varphi}\,\ln\left(
{\sqrt{\frac{\mbox {$2\,U$}}{\mbox {$\sigma_{lg}$}}}}\,
\frac{1}{\cos\varphi}\,\frac{t_{\varphi}}{t}\right)
\,\,\,,\,\,\,t\,=\,\frac{T_w\,-\,T}{T_w}\,\,\,,
\end{eqnarray} 
and
\begin{eqnarray}
l_0\,=\,l_{\infty}\,+\,\frac{\xi}{\sin\varphi}
\,\ln\frac{t}{t\,-\,t_{\varphi}}\,\,\,,
\end{eqnarray}
which diverges for $t\searrow t_{\varphi}$ where
\begin{eqnarray}
t_{\varphi}\,=\,\frac{T_w\,-\,T_{\varphi}}{T_w}\,=\,
\frac{\sqrt{2\,\sigma_{lg}\,U}}{|W_0|}\,\cos\varphi.
\end{eqnarray}
The contact angle is given by
\begin{eqnarray}
\cos\Theta\,=\,1\,+\,\frac{\omega\,(l_{\infty}\sin\varphi)}
{\sigma_{lg}}\,\,\,.
\end{eqnarray}
which leads to (see Eq.(3.10))
\begin{eqnarray}
1\,-\,\cos\Theta\,=\,\frac{1}{2}\,\left(\frac{t}{t_{\varphi}}\right)^2
\,{\cos}^{2}\varphi\,=\,\left(\frac{t}{t_{\varphi}}\right)^2\,v(\varphi)
\,\,\,
\end{eqnarray}
which is in accordance with Eq.(3.11) so that
\begin{eqnarray}
\Theta(T_{\varphi})\,=\,\left(\frac{1}{2}\,\pi\,-\,\varphi\right)
\,\left[1\,-\,\frac{1}{8}\,\left(\frac{1}{2}\,\pi\,-\,\varphi\right)^2
\,+\,O\left(\left(\frac{1}{2}\,\pi\,-\,\varphi\right)^4\right)\right]
\,\,\,.
\end{eqnarray}

For $x\,\geq\,0$ the profile is determined by (see Eq.(3.9))
\begin{eqnarray}
\sin\varphi\,\frac{d\,l(x)}{d\,x}\,=\,
\sqrt{\frac{2\,U}{\sigma_{lg}}}\,\left(\frac{W}{2\,U}\,
+\,\exp\left[-\,\frac{\mbox {$l(x)$}}{\mbox {$\xi$}}
\,\sin\varphi\right]\right)
\end{eqnarray}
with the boundary condition $\left.\frac{\mbox {$d\,l(x)$}}
{\mbox {$d\,x$}}\right|_{x=0^{+}}=\,-\cot\varphi$.
The explicit solution of Eq.(4.8) can be written as
\begin{eqnarray}
l(x)\,=\,l_{\infty}\,+\,\frac{\xi}{\sin\varphi}\,
\ln\left(1\,+\,\frac{1}{\frac{\mbox {$t$}}{\mbox {$t_{\varphi}$}}
\,-\,1}
\exp\left[-\,\frac{t}{t_{\varphi}}\,
\frac{x}{\xi}\,\cos\varphi
\right]\right)\,\,\,,\,\,\,x\,\geq\,0\,\,\,.
\end{eqnarray}
Equation (4.9) is in accordance with all expected limiting behaviors:
$l(0)=l_0$ as given in Eq.(4.3); $t_{\varphi}\rightarrow 0$
for $\varphi\rightarrow \frac{1}{2}\pi$ 
(Eq.(4.4)) so that for
$x$ and $t$ fixed $l(x)\rightarrow l_{\infty}$ which itself
reduces to the planar value $l_{\infty}^{(p)}=\xi\ln\left(
\frac{\mbox {$2U$}}{\mbox {$|W_0|t$}}\right)$ (Eq.(4.2));
and for large $x$ the film thickness $l(x)$ approaches its 
asymptote exponentially from above: 
\begin{eqnarray}
l(x\,\rightarrow\,\infty)\,=\,l_{\infty}\,+\,\frac{\xi}{\sin\varphi}\,
\frac{1}{\frac{\mbox {$t$}}{\mbox {$t_{\varphi}$}}\,-\,1}
e^{-\,\frac{\mbox {$t$}}{\mbox {$t_{\varphi}$}}\,
\frac{\mbox {$x$}}{\mbox {$\xi$}}\,{\mbox {$\cos\varphi$}}}
\,\,\,,
\end{eqnarray}
provided $T$ is not too close to $T_{\varphi}$, i.e.,
$\frac{\mbox {$t$}}{\mbox {$t_{\varphi}$}}-1>>
\exp\left(-\,\frac{\mbox {$t$}}{\mbox {$t_{\varphi}$}}\,
\frac{\mbox {$x$}}{\mbox {$\xi$}}\,{\mbox {$\cos\varphi$}}\right)$.
For any fixed value of $x$ the profile diverges for
$t\rightarrow t_{\varphi}$ as (see Eq.(4.3))
\begin{eqnarray}
l(x\,,\,t\,\rightarrow\,t_{\varphi})\,=\,l_0\,-\,
\frac{\xi}{\sin\varphi}\,\frac{x}{\xi}\,\cos\varphi\,\,\,.
\end{eqnarray}
The maximum curvature, i.e., $l'''(x_0)\,=\,0$ occurs at
\begin{eqnarray}
x_0\,=\,\pm\,\frac{\mbox {$\xi$}}{\frac{\mbox {$t$}}
{\mbox {$t_{\varphi}$}}\,{\mbox {$\cos\varphi$}}}\,
\ln\left(\frac{\mbox {$1$}}{\frac{\mbox {$t$}}{\mbox {$t_{\varphi}$}}
\,-1}\right)\;\;\;\;\;\;\;\;\;\;\;\;\;\;\;\;\;\;\;\;\;\;\;
\;\;\;\;\;\;\;\;\;\;\;\;\;\; \nonumber\\
\;\;\;\;\;\;\;\;\;\;\;\;\;=\,\pm\,\left(1\,-
\,\frac{t\,-\,t_\varphi}
{t_{\varphi}}\right)\,(l_0\,-\,l_{\infty})
\,\tan\varphi\,+\,O(t\,-\,t_{\varphi})
\,\,\,.
\end{eqnarray}
Thus for $t\rightarrow t_{\varphi}$ the position of 
the maximum curvature is given by the intersection of  
the asymptote $f(x)=l_{\infty}+|x|\cot\varphi$
and the horizontal $z=l_0$. Figure 10 illustrates
the change of the meniscus shape upon approaching the filling 
transition temperature $t_{\varphi}$.

The excess coverage $\Gamma$ (see Fig.10) associated with the
meniscus is given by
\begin{eqnarray}
\Gamma\,=\,2\,\Delta\,\rho\,\int\limits_{\mbox {$0$}}
^{\mbox {$\infty$}}\,dx\,[l(x)\,-\,l_{\infty}]\,=\,
\frac{4\,\Delta\,\rho\,\xi^2}{\sin2\,\varphi}\,\frac{t_{\varphi}}{t}
\,I\left(\frac{\mbox {$1$}}{\frac{\mbox {$t$}}{\mbox {$t_{\varphi}$}}
\,-\,1}\right)
\end{eqnarray}
with [63] $I(y)=\int\limits_{0}^{y}
dx\,x^{-1}\ln(1+x)$ (see Fig.11). Since
$I(y\rightarrow \infty)=\frac{1}{6}\pi^2+
\frac{1}{2}\ln^2y$ one finds that upon approaching
the filling transition the excess coverage diverges as
\begin{eqnarray}
\Gamma(t\,\rightarrow\,t_{\varphi})\,=\,
\frac{2\,\Delta\,\rho\,\xi^2}{\sin 2\,\varphi}\,
\ln^2\left(\frac{\mbox {$t$}}{\mbox {$t_{\varphi}$}}\,-\,1\right)
\,\,\,.
\end{eqnarray}
For fixed temperature $\Gamma\left(\varphi\rightarrow
\frac{1}{2}\pi\right)$ vanishes as $\frac{1}{2}\pi-\varphi$
(see Eq.(4.4) and $I(y\rightarrow 0)=y)$. 

The line tension associated with the meniscus shape is given by
Eqs.(3.13), (3.14), (3.2), (3.17), and (4.2) - (4.4). One finds
\begin{eqnarray}
\eta\,=\,-\,2\,\xi\,\sigma_{lg}\,\left[1\,+\,\left(\frac{t}
{t_{\varphi}}\,-\,1\right)\,\ln\left(1\,-\,\frac{t_{\varphi}}
{t}\right)\right]\,\cot\varphi\,\,\,.
\end{eqnarray}
The line tension is negative and approaches its minimal value
at $t=t_{\varphi}$ with a logarithmic singularity (see Fig.12) :
\begin{eqnarray}
\eta(t\,\rightarrow\,t_{\varphi})\,-\,\eta(t\,=\,t_{\varphi})
\,\sim\,-\,\left(\frac{t}
{t_{\varphi}}\,-\,1\right)\,\ln\left(\frac{t}
{t_{\varphi}}\,-\,1\right)\,\,\,.
\end{eqnarray}
For $\varphi \rightarrow \frac{1}{2}\pi$ 
(i.e., $t/t_{\varphi}\rightarrow \infty$) the line tension 
vanishes as
\begin{eqnarray}
\eta\left(\frac{\mbox {$t$}}{\mbox {$t_{\varphi}$}}
\,\rightarrow\,\infty \right)
\,=\,-\,\xi\,\sigma_{lg}\,\left(\frac{t}{t_{\varphi}}
\right)^{-1}\,\cot\varphi\,\,\,.
\end{eqnarray}
Thus for fixed temperature $\eta\left(\varphi\rightarrow
\frac{1}{2}\pi\right)$ vanishes  
$\sim\left(\frac{1}{2}\pi-\varphi\right)^2$ (see Eq.(4.4)).

Finally it should be pointed out that in terms of the variable
$t/t_{\varphi}$ both the excess
coverage $\Gamma$ and the line
tension $\eta$ (see Eq.(4.15) and Fig.12)
can be expressed by scaling functions which
are independent of $\varphi$ and the model parameters
$\sigma_{lg}$, $W_{0}$, and $U$. It will be interesting to
see to which extent this feature is established by more
realistic models.\\

\centerline{\bf V. {Derivation of the effective interface
Hamiltonian}}
\renewcommand{\theequation}{5.\arabic{equation}} 
\setcounter{equation}{0}

The results presented in Secs. III and IV are based on
the effective interface Hamiltonian ${\cal H}[f]$ given
by Eq.(3.1). Although this expression for ${\cal H}[f]$
is rather plausible the status of this equation is that
of a phenomenological ansatz. In this section we describe 
the derivation of Eq.(3.1) from a more basic model in
order to justify our choice of ${\cal H}[f]$ and to gain
insight into the limitations of this form of ${\cal H}[f]$.

For an actual fluid confined to a wedge the appropriate approach
is to start from a density functional theory for the inhomogenous
number density distribution incorporating the full substrate 
potential of the wedge and the dispersion forces between
the fluid particles. However, here we refrain from this
very demanding task of bridging the gap between a fully
microscopic description on an atomic scale and ${\cal H}[f]$.
Instead, by aiming at conceptual insight we focus on the
less ambitious goal to derive ${\cal H}[f]$ from a suitably 
chosen Landau-Ginzburg-Wilson theory for an order parameter
$m$ which corresponds to the deviation of the mean local
number density $\rho({\mbox{\boldmath $r$}})$ of the fluid 
particles from 
the average bulk densities, i.e., $\rho({\mbox{\boldmath $r$}})=
\frac{1}{2}\left(\rho_l+\rho_g\right)+m({\mbox{\boldmath $r$}})$ 
such that $m<0$ corresponds to a gaslike configuration
and $m > 0$ to a liquidlike configuration. The values $m < 0$ $(>0)$ 
are favored by the conjugate field $h < 0$ $(>0)$.   
This bulk field is proportional to $\Delta \mu=\mu_0(T)-\mu$
such that for $h>0$ there is the liquid phase whereas for
$h<0$ one has the gas phase. For the present wedge geometry
(see Fig.1) the natural form of the {\it L}andau-{\it G}inzburg
-{\it W}ilson Hamiltonian is [64] 
\begin{eqnarray}
{\cal H}_{LGW}[m]\,=
\int\limits_{{\mbox {$-\,\infty$}}}
^{{\mbox {$\infty$}}}dx\left\{\int 
\limits_{{\mbox {$|x|\cot\varphi$}}}
^{{\mbox {$\infty$}}} dz\,\left(\frac{K}{2}
\left[\left(\frac{\partial m(x\,,\,z)}
{\partial z}\right)^2\,+\,\left(\frac
{\partial m(x\,,\,z)}{\partial x} 
\right)^2\right]\,+\,\right. \right.\nonumber \\ \left. \left. 
\Phi\,(m(x\,,\,z)\right)
\,+\,\frac{1}{\sin\varphi}\,\Phi _{1}(m(x), 
z=|x|\cot\varphi)\right\}\,\,\,.
\end{eqnarray}
Here we have already assumed translational invariance in the $y$
direction. $\Phi(m)=\Phi_0(m)-mh$ denotes the bulk
free energy density. The bulk free energy density at coexistence
$h=0$, i.e., $\Phi_0$ exhibits two equally deep minima
located at $m_{\alpha 0}<0$ and $m_{\beta 0}>0$
corresponding to the gas ($\alpha$) and liquid ($\beta$)
densities, respectively. With the bulk phase being vapor
we have either $h=0^{-}$ at coexistence or $h<0$
off coexistence. In the present context we employ the 
so-called double-parabola model which is the simplest form of
$\Phi_0(m)$ which allows one to obtain analytical results
[59,65-67]:
\begin{eqnarray}
\Phi _{0}(m)\,=\,\left\{
\begin{array}{ccl}
\frac{1}{2}K\xi_{\alpha}^{-2}(m\,-\,m_{\alpha 0})^2 &\,\,\,,
\,\,\,m\,\,\,\leq\,0\,\,\,\,\\\\
\frac{1}{2}K\xi_{\beta}^{-2}(m\,-\,m_{\beta 0})^2 &\,\,\,,
\,\,\,m\,\,\,\geq\,0\,\,\,.\\
\end{array}
\right.
\end{eqnarray}
$\xi_{\gamma}$ denotes the correlation length in the bulk phase
$\gamma\,=\,\alpha\,\,,\,\,\beta$. The parameters are chosen such 
that $\Phi_0(m)$ is continuous at $m\,=\,0$, i.e.,  
$\xi_{\alpha}^{-1}m_{\alpha 0}=-\xi_{\beta}^{-1}m_{\beta 0}$.
The term $-mh$ in $\Phi(m)$ can be absorbed into the double-parabola form 
as follows:
\begin{eqnarray}
\Phi\,(m)\,=\,\left\{
\begin{array}{ccl}
\frac{1}{2}K\xi_{\alpha}^{-2}(m\,-\,m_{\alpha h})^2\,+\,A_{\alpha} 
&\,\,\,,\,\,\,m\,\,\,\leq\,0\,\,\,\,\\
\\
\frac{1}{2}K\xi_{\beta}^{-2}(m\,-\,m_{\beta h})^2\,+\,A_{\beta} 
&\,\,\,,\,\,\,m\,\,\,\geq\,0\,\,\,\,,\\
\end{array}
\right.
\end{eqnarray}
where $m_{\gamma h}=m_{\gamma 0}+h\xi_{\gamma}^{2}/K$ and
$A_{\gamma}=-m_{\gamma 0}h-h^2\xi_{\gamma}^{2}/(2K)$ is the
equilibrium bulk free energy density of the $\gamma$ phase.
Off coexistence, i.e., for $h<0$ the $\alpha$ phase is the stable 
phase with $m_{\alpha h}=m_{\alpha 0}+
h\xi_{\alpha}^{2}/K$ as the equilibrium value 
of the order parameter. The requirement of the occurrence of 
the metastable liquid phase, i.e., $m_{\beta h}>0$ leads to the
constraint $h>-K\xi_{\beta}^{-2}m_{\beta 0}$.

For the surface contribution in Eq.(5.1) we choose the common
[64] expression
\begin{eqnarray}
\Phi_{1}(m_1)\,=\,-h_1\,m_1\,-\,\frac{1}{2}\,g\,m_1
\end{eqnarray}
where $m_1(x)\,=\,m(x\,,\,z\,=\,|x|\,\cot\varphi)$. The surface
field $h_1$ is taken to be positive so that it has the opposite
sign as $h$. It favors the liquid phase $\beta$ near the wall
 whereas the bulk field $h$ favors the vapor phase $\alpha$.
The surface enchancement parameter $g$ is taken to be negative. In
Eq.(5.1) the prefactor $\frac{\mbox {$1$}}{\mbox {$\sin\varphi$}}$
of the surface term
takes into account the increase of the surface area as compared to 
its horizontal projection. In general the LGW-Hamiltonian can
additionally contain a line contribution corresponding to the
edge of the wedge as well as a modified surface term including
lateral derivatives of the order parameter. Since it turns out that
these contributions and modifications do not influence the filling
transition temperature of the wedge we do not consider them here.

Following the approach of Fisher and Jin [59,66-68] we employ the
so-called crossing criterion in order to construct the 
effective interface Hamiltonian corresponding to the type of interface 
configurations as depicted in Fig.1(b). Order parameter configurations
$m(x\,,\,z)$, which are monotonic and compatible with 
the boundary conditions to be positive near the walls and negative 
in the bulk, exhibit a line $z=f(x)$ along which $m(x,z)$
is zero. The crossing criterion identifies this line as the
position of the interface:
\begin{eqnarray}
m(x\,,\,z\,=\,f(x))\,=\,0\,\,\,.
\end{eqnarray}
The construction scheme for obtaining the effective interface
Hamiltonian is to minimize Eq.(5.1) under the constraint of the boundary
conditions and under the constraint given by Eq.(5.5) for a prescribed
configuration $f(x)$.
The resulting order parameter ${\overline m}(x,z;[f])$, 
which corresponds to this minimum and thus is a functional of
$f(x)$, is inserted into Eq.(5.1) and yields as the surface contribution
to ${\cal H}_{LGW}$ the effective interface Hamiltonian ${\cal H}[f]$.
The equilibrium shape ${\bar f}(x)$ of the meniscus minimizes ${\cal H}[f]$.
(In the following we drop overbars.)
$\frac{\mbox {$\delta {\cal H}_{LGW}[m]$}}{\mbox
{$\delta m(x,z)$}}=0$ yields (see Eq.(5.1))
\begin{eqnarray}
K\,\left(\frac{\partial ^2}{\partial x^2}\,+\,
\frac{\partial ^2}{\partial z^2}\right)m_{\gamma}(x\,,\,z)\,
=\,\Phi'(m_{\gamma}(x\,,\,z))\,\,\,,
\end{eqnarray}
and the boundary condition at the substrate
\begin{eqnarray}
K\left.\left(\frac{\partial m_{\beta}}{\partial z}\,-\,
{\rm sgn}(x)\,\cot\,\varphi\,\,\,\frac{\partial m_{\beta}}
{\partial x}\right)\right|_{z=|x|\cot \varphi}\,=\,
\frac{1}{\sin\varphi}\,\left.\Phi_1'(m_{\beta})
\right|_{z=|x|\cot \varphi} \;\;\nonumber\\
\;\;=\,-\,\left.
\frac{h_{1}\,+\,gm_{\beta}}{\sin\varphi}
\right|_{z=|x|\cot \varphi}
\,\,\,
\end{eqnarray}
supplemented by the bulk boundary condition
\begin{eqnarray}
\lim_{z \rightarrow \infty} m_{\alpha}(x\,,\,z)\,
=\,m_{\alpha h}\,\,\,.
\end{eqnarray}
The indices $\alpha$, $\beta$, and $\gamma=\alpha$, $\beta$
indice that in the special case of the double-parabola model,
which will be considered henceforth, either the upper or lower
part of Eq.(5.3) must be used depending on the sign of $m$; 
$\alpha$ corresponds to $m<0$ and $\beta$ corresponds to
$m>0$. In Eq.(5.7) we have used the index $\beta$ assuming
that $m$ is positive near the wall; otherwise the concept of an
interface description would not be applicable.

Following Refs.[66] and [67] the solution of Eqs.(5.5)-(5.8) 
is expanded into a series of terms which contribute
increasing orders of derivatives of 
$l(x)=f(x)-|x|\cot\varphi$ :
\begin{eqnarray}
m_{\gamma}\,=\,m_{\gamma \pi}\,+\,m_{\gamma 1}\,+\,m_{\gamma 2}\,+\,
...\,
\end{eqnarray} 
where the index $\pi$ indicates the planar limit $2\varphi=
\pi$. For the individual terms one obtains within the 
double-parabola model (Eq.(5.3))
\begin{eqnarray}
m_{\gamma \pi}\,=\,m_{\gamma 0}\,+\,F_{\gamma 0}
\left(z\,-\,|x|\cot \varphi,l(x)\right)\,\,\,,\\
m_{\gamma 1}\,=\,F_{\gamma 1} \left(z\,-\,|x|\cot \varphi,l(x)\right)
\frac{dl(x)}{dx}\;,\;\;\;\;
\end{eqnarray}
and
\begin{eqnarray}
m_{\gamma 2}\,=\,F_{\gamma 21}
\left(z-|x|\cot \varphi , \,l(x)\right)\left(\frac{dl(x)}{dx}\right)^2
+F_{\gamma 22}\left(z-|x|\cot \varphi ,\, l(x)\right)
\frac{d^2 l(x)}{dx^2}.
\end{eqnarray}
The functions $F_{\gamma i}$ and $F_{\gamma 2i}$ depend on the lateral 
coordinate $x$ only via the vertical distance $z-|x|\cot\varphi$ from
the walls and via the local film thickness $l(x)$ but not separately.
We note that due to the cusp nonanalyticity of $l(x)$
at $x=0$ the higher 
order terms in Eq.(5.9) contain distributions which are singular
at $x=0$ (see, e.g., Eq.(5.12)). Since $m(x,z)$ is smooth
for $z\neq f(x)$ (see Eqs.(5.3) and (5.5)) these singularities
have to cancel each other in Eq.(5.9). Since in the following we
focus on the lowest terms we do not pursue this aspect further.

The function $F_{\gamma 0}(u,v)$ satisfies the differential
equation
\begin{eqnarray}
\left(\frac{1}{\sin^2\varphi}\,\frac{\partial^2}{\partial\,u^2}
\,-\,\xi_{\gamma}^{-2}\right)\,F_{\gamma 0}(u\,,\,v)\,=\,0
\end{eqnarray}
with the boundary condition
\begin{eqnarray}
\left.K\,\frac{\partial}{\partial\,u}\,
F_{\beta 0}(u\,,\,l(x))\right|_{u\,=\,0}\,=\,
-\,(h_1\,+\,g\,m_{\beta 0}\,+\,g\,F_{\beta 0}(u\,=\,0\,,\,
l(x)))\,\sin\varphi\,\,\,.
\end{eqnarray}
Equations (5.13) and (5.14) together with Eq.(5.8) lead to the
first contribution $m_{\gamma \pi}$ in Eq.(5.9) which does not
depend on derivatives of $l(x)$:
\begin{eqnarray}
m_{\alpha \pi}(x\,,\,z\,;\,[f])\,=\,m_{\alpha 0}\,\left[
1\,-\,\exp\left(-\,\frac{z\,-\,f(x)}{\xi_{\alpha}}\,\sin\varphi
\right)\right]
\end{eqnarray}
and
\begin{eqnarray}
m_{\beta \pi}(x\,,\,z\,;\,[f])\,=\,m_{\beta 0}\,+\,
B_{+}\,\exp\left(\frac{z\,-\,f(x)}
{\xi_{\beta}}\,\sin\varphi\right)
\,+\, \nonumber \\ B_{-}\,\exp\left(-\,\frac{z\,-\,f(x)}{\xi_{\beta}}\,
\sin\varphi\right)
\end{eqnarray}
where the coefficients $B_{\pm}$ are given as
\begin{eqnarray}
B_{+}\,=\,-\,\frac{m_{\beta 0}\,+\,\tau\,X}
{1\,-\,{\cal G}\,X^2}
\end{eqnarray}
and
\begin{eqnarray}
B_{-}\,=\,\frac{\tau\,+\,{\cal G}\,m_{\beta 0}\,X}
{1\,-\,{\cal G}\,X^2}
\end{eqnarray}
with
\begin{eqnarray}
X\,=\,\exp\left(-\,\frac{l(x)}{\xi_{\beta}}\,\sin\varphi\right)
\,\,\,,\\
\tau\,=\,\frac{h_{1}\,+\,g\,m_{\beta 0}}
{K\,\xi_{\beta}^{-1}\,-\,g}\,\,\,,\;\;\;\;\;\;\;\;\;\;\;\;
\end{eqnarray}
and
\begin{eqnarray}
{\cal G}\,=\,-\,\frac{K\,\xi_{\beta}^{-1}\,+\,g}
{K\,\xi_{\beta}^{-1}\,-\,g}\,\,\,.
\end{eqnarray}
The remaining functions $F_{\gamma 1}(u,v)$,
$F_{\gamma 21}(u,v)$, and $F_{\gamma 22}(u,v)$
satisfy the following equations:
\begin{eqnarray}
\left(\frac{1}{\sin^2\varphi}\,\frac{\partial^2}{\partial u^2}
\,-\,\xi_{\gamma}^{-2}\right)\,F_{\gamma 1}(u\,,\,v)\,=\,
2\,\cot\varphi\,\frac{\partial^2 F_{\gamma 0}(u\,,\,v)}
{\partial u\,\partial v}\,\,\,,
\;\;\;\;\;\;\;\;\;\;\;\;\;\;\;\;\;\;\;\;\;\;\\
\left(\frac{1}{\sin^2\varphi}\,\frac{\partial^2}{\partial u^2}
\,-\,\xi_{\gamma}^{-2}\right)\,F_{\gamma 21}(u,\,v)=
2\cot\varphi \frac{\partial^2 F_{\gamma 1}(u,\,v)}
{\partial u\,\partial v}
\,-\,\frac{\partial^2 F_{\gamma 0}(u\,,\,v)}{\partial v^2}\,\,\,,
\end{eqnarray}
and
\begin{eqnarray}
\left(\frac{1}{\sin^2\varphi}\,\frac{\partial^2}{\partial u^2}
\,-\,\xi_{\gamma}^{-2}\right)\,F_{\gamma 22}(u\,,\,v)\,=\,
2\,\cot\varphi\,\frac{\partial F_{\gamma 1}(u\,,\,v)}{\partial u}
\,-\,\frac{\partial F_{\gamma 0}(u\,,\,v)}{\partial v}\,\,\,.
\end{eqnarray}
For constructing the effective Hamiltonian up to
square-gradient terms only the functions $F_{\gamma 0}$ 
and $F_{\gamma 1}$ are needed. The solution of Eq.(5.22) is
proportional to $\cot\varphi$ and only the square of 
$F_{\gamma 1}$ contributes to the effective Hamiltonian. The terms 
proportional to $F_{\gamma i}$, $i=1, 21, 22$, vanish because these 
terms are linear in deviations from $m_{\gamma \pi}$ which itself 
minimizes the flat substrate Hamiltonian; thus the prefactors 
multiplying these deviations vanish. 
For a wedge with a wide opening angle, i.e., $\varphi$ 
close to $\frac{1}{2}\pi$, one has $\cot\varphi<<1$
so that in the spirit, which led to Eq.(3.7), this contribution
can be neglected here, too. Thus, the effective Hamiltonian is
obtained by inserting $m_{\gamma \pi}$ as given by Eqs.(5.14)-(5.20)
 into the LGW-Hamiltonian (Eq.(5.1)). For a wide
wedge only terms up to $\left(\frac{\mbox {$dl(x)$}}
{\mbox {$dx$}}\right)^2$ are retained leading $-$ after subtracting the 
surface contribution to the free energy $-$ to 
\begin{eqnarray}
{\cal H}[l]\,=\,\int\limits_{-\infty}^{\infty}\,
dx\left\{\frac{\sin \varphi}{2}
\Sigma_{\alpha \beta}(l\sin\varphi)\,\left(\frac{dl(x)}{dx}\right)^2\,
+\,\chi_{\alpha\beta} (l(x)\sin \varphi)\,\frac{dl(x)}{dx}\cos\varphi
\right.\nonumber\\
\left.
+\,\frac{V(l(x)\,\sin\varphi)\,-\, V(l_{\infty}\,\sin\varphi)}{\sin \varphi}
\right\}\,\,\,.
\end{eqnarray}
The coefficients $\Sigma_{\alpha \beta}$ and $\chi_{\alpha \beta}$
and the potential $V$ are given by
\begin{eqnarray}
\Sigma_{\alpha \beta}(l(x)\,\sin\varphi)\,=\,\frac{K}{\sin\varphi}\,
\left[\,\int\limits_{\mbox {$|x|\,\cot\varphi$}}^{\mbox {$f(x)$}}
dz\,\left(\frac{\partial m_{\beta \pi}}{\partial l}\right)^2
\,+\,\int\limits_{\mbox {$f(x)$}}^{\mbox {$\infty$}}
dz\,\left(\frac{\partial m_{\alpha \pi}}{\partial l}\right)^2
\right]\,\,\,,\\
\chi_{\alpha \beta}(l(x)\,\sin\varphi)\,=\,-\,\frac{K}{\sin\varphi}\,
\left[\,\int\limits_{\mbox {$|x|\,\cot\varphi$}}^{\mbox {$f(x)$}}
dz\,\frac{\partial m_{\beta \pi}}{\partial z}
\,\frac{\partial m_{\beta \pi}}{\partial l}\,+\,
\int\limits_{\mbox {$f(x)$}}^{\mbox {$\infty$}}
dz\,\frac{\partial m_{\alpha \pi}}{\partial z}\,
\frac{\partial m_{\alpha \pi}}{\partial l}\right]
\,\,\,,
\end{eqnarray}
and
\begin{eqnarray}
V(l(x)\,\sin\varphi)\,=\,\;\;\;\;\;\;\;\;\;\;\;\;\;\;\;\;\;
\;\;\;\;\;\;\;\;\;\;\;\nonumber\\
\sin\varphi\,
\int\limits_{\mbox {$|x|\,\cot\varphi$}}^{\mbox {$f(x)$}}
dz\,\left[\Phi({m_\beta \pi})\,-\,\Phi({m_{\alpha h}})\,+\,
\frac{K}{2\,\sin^2\varphi}\left(\frac{\partial m_{\beta \pi}}
{\partial z}\right)^2\right] \\
+\,\sin\varphi\,
\int\limits_{\mbox {$f(x)$}}^{\mbox {$\infty$}}
dz\,\left[\Phi(m_{\alpha \pi})\,-\,\Phi({m_{\alpha h}})\,+\,
\frac{K}{2\,\sin^2\varphi}\left(\frac{\partial m_{\alpha \pi}}
{\partial z}\right)^2\right] \nonumber\\
+\,\Phi_1(m_{\beta \pi}(x\,,\,z\,=\,|x|\,\cot\varphi\,;\,[f]))
\,\,\,.\;\;\;\;\;\;\;\;\;\;\;\;\;\;\;\;\;\;\;\;\;\;\;\;\;
\;\;\;\;\;\;\;\;\;\;\;\;\;\;\nonumber
\end{eqnarray}
In Eqs.(5.26)-(5.28) the profiles $m_{\alpha \pi}$ and
$m_{\beta \pi}$ are given by Eqs.(5.15) and (5.16), respectively,
and $m_{\alpha h}=m_{\alpha 0}+h\xi_{\alpha}^2/K$
is the bulk value of the order parameter. The differentiation with
respect to $l$ is a differentiation with respect to $l(x)$ for
any fixed value of $x$. The stiffness coefficients
$\Sigma_{\alpha \beta}$ and $\chi_{\alpha \beta}$ 
as well as the potential $V$ turn out to have the same
functional dependence on the thickness of the wetting film as for the
corresponding planar substrate.  Here they are evaluated 
at the local normal distance $l(x)\sin\varphi$ to the wall (see
Fig.1(b)). For large thicknesses of the wetting film both
stiffness coefficients $\Sigma_{\alpha \beta}$ and
$\chi_{\alpha \beta}$ approach constant values which are the surface
tension $\sigma_{\alpha \beta}$ in both cases. The term linear in
$\frac{\mbox {$dl(x)$}}{\mbox {$dx$}}$, which is known from
studies of wetting on corrugated substrates [67], is multiplied 
by $\cos\varphi$ and thus for a wide wedge it can be neglected
as compared with the other two contributions in Eq.(5.25). Therefore
we can conclude that the phenomenological ansatz for the effective
Hamiltonian in Eq.(3.7) can be derived systematically from
a more basic theory in the limit of a wide opening angle of
the wedge. The derivation shows which kind of terms are left
out by the ansatz in Eq.(3.7).

Moreover, the model parameters entering into Eq.(3.7) can be
expressed in terms of those of the underlying LGW-Hamiltonian.
Within the double-parabola model and the identifications
$\alpha=g$ and $\beta=l$ one finds
\begin{eqnarray}
\sigma_{\alpha \beta}\,=\,\frac{1}{2}\,K\,(
\xi_{\beta}^{-1}\,m_{\beta 0}^2\,+\,\xi_{\alpha}^{-1}\,m_{\alpha 0}^2)
\,\,\,,\;\;\;\;\;\;\;\;\;\;\;\;\;\;\;\;\;\;\;\;\;\;\;\;\;\;\;\;\;\;\;\,\\
\sigma_{w \beta}\,=\,\frac{1}{2}\,K\,\xi_{\beta}^{-1}\tau^2
\,-\,h_{1}\,(\tau\,+\,m_{\beta 0})
\,-\,\frac{1}{2}\,g\,(\tau\,+\,m_{\beta 0})^2
\,\,\,,
\end{eqnarray}
and
\begin{eqnarray}
\omega(L)\,=\,\frac{W\,\exp\left(-\,
\frac{\mbox {$L$}}{\mbox {$\xi_{\beta}$}}\right)\,
+\,U\,\exp\left(\frac{\mbox {$2\,L$}}{\mbox {$\xi_{\beta}$}}\right)}
{1\,-\,{\cal G}\,\exp\left(\frac{\mbox {$-2\,L$}}{\mbox {$\xi_{\beta}$}}
\right)}\;\;\;\;\;\;\;\;\;\;\;\;\;\;\;\;\;\;\;\;\;\;\;\;\;\;
\;\;\;\;\;\;\;\;\;\;\;\nonumber\\
=\,W\,\exp\left(-\,\frac{\mbox {$L$}}{\mbox {$\xi_{\beta}$}}\right)\,
+\,U\,\exp\left(-\,\frac{\mbox {$2\,L$}}{\mbox {$\xi_{\beta}$}}\right)\,+\,
O\left(\exp\left(-\,\frac{\mbox {$3\,L$}}{\mbox {$\xi_{\beta}$}}\right)\right)
\end{eqnarray}
with
\begin{eqnarray}
W\,=\,2\,K\,\xi_{\beta}^{-1}m_{\beta 0}\,\tau
\end{eqnarray}
and
\begin{eqnarray}
U\,=\,K\,\xi_{\beta}^{-1}\,({\cal G}\,m_{\beta 0}^2
\,+\,\tau^2)
\end{eqnarray}
(compare Eqs.(3.2) and (3.17)). Acccording to Eqs.(5.20) and
(5.32) $W$ is negative at low temperatures, at which 
$m_{\beta 0}>0$ is large, and vanishes at $\tau=0$
for $m_{\beta 0}=h_1/|g|$ which is an implicit
equation for the wetting transition temperature.
For $|g|>K/\xi_{\beta}$ one has 
$U(\tau=0)>0$ and the model exhibits a critical
wetting transition as studied in Sec.IV. Thus if in Eq.(5.31)
$\omega(L)$ is truncated after the first two terms we 
fully recover the model analyzed in the previous section.

Finally we note that in the limit of a wedge with wide
opening angle the resulting form of ${\cal H}[f]$ in Eq.(5.24)
does not depend on the special choice of $\Phi(m)$
corresponding to the double-parabola model, i.e., 
Eq.(5.3) but holds for a general functional form $\Phi(m)$,
which describes two-phase coexistence with a critical point.
In that case Eq.(5.13) is replaced by
\begin{eqnarray}
\frac{\partial^2}{\partial u^2}\,F_{\gamma 0}(u\,,\,v)\,=\,
\Phi'(F_{\gamma 0}(u\,,\,v))\,\sin^2\varphi\,\,\,.
\end{eqnarray}
Moreover the fact, that in Eq.(5.3) the functions 
$\Sigma_{\alpha \beta}$, $\chi_{\alpha \beta}$, and
$V$ exhibit the same functional form as for the planar
substrate, holds also for a general expression for $\Phi(m)$.\\

\centerline{\bf VI. {Summary}}
\renewcommand{\theequation}{4.\arabic{equation}} 

We have obtained the following main results for the structure
of a fluid exposed to a substrate forming a wedge with opening
angle $2\,\varphi$ (Fig.1(a)):
\begin{enumerate}
{\item A nonvolatile liquid spreads along the edge of the wedge
if its contact angle $\Theta$ on the corresponding planar
substrate is less than $\frac{1}{2}\pi-\varphi$
(see Fig.2 and Eq.(2.14)). A vapor bubble in a liquid spreads 
along the wedge if $\Theta>\frac{1}{2}\pi+\varphi$ 
(Eq.(2.20)). Theory of capillarity tells that $\Theta(T=
T_{\varphi})=\frac{1}{2}\pi-\varphi$ marks also
the filling transition temperature $T_\varphi$ of a wedge by 
a volatile liquid in eqilibriun with its vapor reservoir
(Subsec.IIB). At liquid-vapor coexistence of the bulk
phases the wedge is completely filled by the liquid phase
for $T>T_{\varphi}$ although $T_{\varphi}<T_w$ where
$T_w$ with $\Theta(T=T_w)=0$ denotes the wetting
transition temperature of the corresponding planar substrate;
$T_{\varphi\rightarrow \frac{1}{2}\pi}=T_w$.
The filling transition constitutes a nonanaliticity in the
{\it surface contribution} to the free energy of the liquid 
confined by the wedge (Eq.(2.18)).}
{\item For $T<T_{\varphi}$ or off liquid-vapor coexistence
the surfaces of the wedge are covered by a thin wetting film
which requires a more
detailed microscopic description, e.g., by an
effective interface Hamiltonian ${\cal H}[f]$ for the shape
$f(x)$ of the ensuing meniscus of the emerging liquid-vapor
interface (see Eqs.(3.1) and (3.7) and Fig.1(b)).
The interaction of this interface with the substrate is
governed by the effective interface potential $V$ (Eq.(3.2)).
This description allows one to compute the {\it line
contribution} to the free energy which determines the shape
of the meniscus. Without specifying the explicit functional
form of $V$ the dependence of the filling height $l_0$
(Fig.1(b)) on temperature and deviation $\Delta\mu$ from 
two-phase coexistence can be discussed graphically 
(Figs.3 and 4). The filling transition at $T_{\varphi}$ and
$\Delta\mu=0$ can be continuous or discontinuous; the
order of the filling transition is the same as the order of the
wetting transition of the corresponding planar substrate.
Quite generally $l_0$ diverges $\sim (\Delta\,\mu)^{-1}$
upon approaching coexistence, i.e., $\Delta\mu\rightarrow0$,
for $T > T_{\varphi}$ (Eq.(3.12)).}
{\item By analyzing the line tension (Eqs.(3.13) and (3.14))
graphically (Figs.5-7) one finds that a first-order filling
transition at coexistence, at which $l_0$ jumps from a microscopic
value to a macroscopic one, is accompanied by a prefilling line
extending into the vapor phase region of the bulk phase
diagram (Fig.8). This prefilling line is the locus of nonanalyticities
in the line tension; there the surface and bulk contributions to
the free energy are analytic. Upon crossing the prefilling line
the filling height undergoes a first-order thin-thick
transition (Fig.9). The prefilling line joins the line 
$\Delta\mu=0$ of the bulk coexistence tangentially.
For increasing opening angles $T_{\varphi}$ moves towards $T_w$.
Accordingly the prefilling line slides into the prewetting
line and breaks up into two pieces (Figs.8 and 9) giving rise
to rich reentrant prefilling transitions. These general features are 
born out explicitly by model calculations based on a specific
choice of the effective potential (Eq.(3.17)).}
{\item The prefilling transition as obtained from mean-field theory
is smeared out by fluctuations of the local filling height along
the edge of the wedge (Subsec.IIIE). Instead of the jump 
the mean filling height changes smoothly between a small value
$l_0^{<}$ and a large value $l_0^{>}$ near the prefilling line.
For small undersaturations $\Delta\mu$ the larger value 
diverges $\sim (\Delta\mu)^{-1}$ so that in this limit the
temperature resolution required for distinguishing between
the jump and the actual smooth crossover is experimentally not accessible 
(see Eq.(3.19)). Due to $l_0^{>}(\Delta\mu=0)=\infty$ the
filling transition at coexistence persists even in the strict sense.
Thus for a system, which exhibits a first-order wetting
transition in planar geometry, the prefilling line in a wedge
should be detectable in experiments.} 
{\item Whereas the thermodynamic behaviour of gross features such
as the filling height $l_0$ and thus the filling transition itself
can be obtained on rather general grounds, the determination of the
actual meniscus shape requires model calculations based on
explicit choices for the interface effective potential $V$.
For short-ranged forces  (see Eq.(3.17) with C=0) 
exhibiting a continuous wetting transition the meniscus shape
(Eq.(4.9) and Fig.10), the excess coverage (Fig.11), and the line
tension (Fig.12) can be obtained analytically. In terms of the
reduced temperature variable $t/t_{\varphi}
=(T_w-T)/(T_w-T_{\varphi})$,
within this model the excess coverage and the line tension are
governed by scaling functions (Figs.11 and 12) which are 
independent of the opening angle $\varphi$ and of potential parameters.
The scaling functions are nonanalytic for $T\rightarrow T_{\varphi}$
(see Eqs.(4.14) and (4.16)).}
{\item In the limit of a  wide opening angle $\varphi$ the effective 
interface Hamiltonian ${\cal H}[f]$ (Eq.(3.7)) can be deduced
from a Landau-Ginzburg-Wilson Hamiltonian (Eq.(5.1)). This
derivation (Sec.V) points towards additional terms in ${\cal H}[f]$ which
appear for smaller opening angles and which are not yet 
included in Eq.(3.7).}
\end{enumerate}
\vspace*{5mm}
{\bf Acknowledgements :}
It is a pleasure for us to acknowledge helpful discussions
and comments by T. Boigs, R. Evans, and T. Getta.\\

\centerline{\bf {Appendix A: Constrained equilibrium in a nonsymmetric 
wedge}}
\renewcommand{\theequation}{A\arabic{equation}}
\setcounter{equation}{0}

The analysis presented in Subsec.IIA can be repeated 
for a nonsymmetric wedge characterized by
contact angles $\Theta_1$ and $\Theta_2$ 
on the left and right side of the wedge, respectively.
Also in this case the liquid with constrained 
volume $V$ forms a single spherical liquid-gas interface or a bridge.
In each case the interface intersects the sides of the wedge at angles
which are equal to the corresponding contact angles.

In the case of a single interface the radius $R$ of the
corresponding circle is 
\begin{eqnarray}
R\,=\,\sqrt{V}\left\{\frac{\Theta_1\,+\,\Theta_2\,+\,2\,\varphi
\,-\,\pi}{2}\,-\,\left(\frac{\cos\Theta_1\,-\,\cos\Theta_2}{2\,\cos\varphi}
\right)^2\,\cot\varphi\,\right.\nonumber\\
\left.+\,\frac{\cos\Theta_1\,\cos(\Theta_1\,+\,\varphi)
\,+\,\cos\Theta_2\,\cos(\Theta_2\,+\,\varphi)}{2\,\sin\varphi}\right\}
^{-\frac{1}{2}}\,\,\,
\end{eqnarray}
while the surface free energy difference between 
the filled and the nonfilled wedge is given by
\begin{eqnarray}
\Delta\,F\,=\,\pm\,2\,\sigma_{lg}\,
\sqrt{V}\left\{\frac{\Theta_1\,+\,\Theta_2\,+\,2\,\varphi
\,-\,\pi}{2}\,-\,\left(\frac{\cos\Theta_1\,-\,\cos\Theta_2}{2\,\cos\varphi}
\right)^2\,\cot\varphi\,\right.\nonumber\\
\left.+\,\frac{\cos\Theta_1\,\cos(\Theta_1\,+\,\varphi)
\,+\,\cos\Theta_2\,\cos(\Theta_2\,+\,\varphi)}{2\,\sin\varphi}\right\}
^{\frac{1}{2}}\,\,\,
\end{eqnarray}
where the upper sign corresponds to a convex and 
the lower sign to a concave
interface. The interface becomes flat and $\Delta F$ vanishes
at a temperature $T_{\varphi}$ determined implicitly by
\begin{eqnarray}
\Theta_1(T_{\varphi})\,+\,\Theta_2(T_{\varphi})\,+\,
2\,\varphi\,=\,\pi\,\,\,.
\end{eqnarray}
This temperature marks a filling phase transition for a nonsymmetric
wedge if the constraint of a fixed volume is removed in favor of 
a grand canonical ensemble. 
At $T=T_{\varphi}$ both the filled and the
nonfilled configuration have the same surface free
energies, independent of the volume of the liquid.

A bridge configuration in a nonsymmetric wedge
is possible as well. The common  radius $R$ of both
interfaces is 
\begin{eqnarray}
R\,=\,\sqrt{V}\left\{\Theta_1\,+\,\Theta_2\,-\,\pi\,-\,
\frac{1}{2}\,\left(\sin 2\,\Theta_1\,+\,\sin 2\,\Theta_2\right)\right\}
^{-\frac{1}{2}}\,\,\,
\end{eqnarray}
while the free energy difference between the liquid bridge 
and the nonfilled wedge is given by
\begin{eqnarray}
\Delta\,F\,=\,4\,\sigma_{lg}\,V\,
\left\{\varphi\,+\,\left[\frac{\cos^2\Theta_1\,+\,\cos^2\Theta_2}{2}\,
-\,2\,\left(\frac{\cos\Theta_1\,-\,\cos\Theta_2}{2\,\cos\varphi}
\right)^2\right]\,\cot\varphi\right\}\nonumber\\
\times\,
\left\{\Theta_1\,+\,\Theta_1\,-\,\pi\,-\,\frac{1}{2}\,
\left(\sin 2\,\Theta_1\,+\,\sin 2\,\Theta_2\right)\right\}
^{-\frac{1}{2}}\,\,\,.
\end{eqnarray}
The bridge configuration occurs provided 
the following condition is fulfilled:
\begin{eqnarray}
\left(\frac{\cos\Theta_1\,+\,\cos\Theta_2}{2\,\sin\varphi}\right)^2\,
+\,\left(\frac{\cos\Theta_1\,-\,\cos\Theta_2}{2\,\cos\varphi}\right)^2
\,>\,1\,\,\,.
\end{eqnarray}
For $\Theta _{1}=\Theta _{2}$ 
these expressions reduce to those given in Sec.II.
For the symmetric wedge
the interface is concave if $\Theta+\varphi<
\frac{1}{2}\pi$ and convex if $\Theta+\varphi>\frac{1}{2}\pi$.
In the nonsymmetric case it is possible 
for an interface to be convex or concave or even form a bridge
if $\Theta_1+\varphi<\frac{1}{2}\pi$ and simultaneously 
$\Theta_2+\varphi>\frac{1}{2}\pi$. \\

\centerline{\bf {Appendix B: Expansions for a wide wedge}}
\renewcommand{\theequation}{B\arabic{equation}}
\setcounter{equation}{0}

For a wide wedge the opening angle is close to $\pi$
so that $\varphi=\frac{1}{2}\pi-\varepsilon$ with $\varepsilon \ll 1$.
In this limit $\frac{\mbox {$df$}}{\mbox {$dx$}} \equiv f_x$ 
is small for all $x$ so that the first part of the
integrand in Eq.(3.1) can be expanded into powers of $f_x^2$.
In terms of $l_x=f_x-\cot\varphi\,\,,\,\,x\geq 0$ one has
\begin{eqnarray}
\sqrt{1+f_x^2}\,-\frac{1}{\sin\varphi}\,=\,\frac{1}{\sin\varphi}\,
\left\{\left(1\,+\,2\,l_x\,\cos\varphi\,\sin\varphi\,+\,
l_x^2\,\sin^2\varphi\right)^{\frac{1}{2}}\,-\,1 \right\}\nonumber\\
=\,\frac{1}{\sin\varphi}\,\left\{\frac{1}{2}\,l_x^2\,\sin^2\varphi
\,+\,l_x\,\sin\varphi\,\cos\varphi\,+\,O(\varepsilon ^4) \right\}
\;\;\;\;\\
=\,\frac{1}{\sin\varphi}\,\left\{\frac{1}{2}
f_x^2\,\sin^2\varphi-\,\frac{1}{2}\,
\cos^2\varphi\,+\,O(\varepsilon ^4) \right\}\;\;\;\;\;\;\;\;\,
\;\;\;\nonumber\\=\,
\frac{1}{2}\,\sin\varphi\,\left\{f_x^2\,-\,\cot^2\varphi\right\}
\,+\,O(\varepsilon ^4)\;\;\;\;\;\;\;\;\;\;\;\;\;\;\;\;\;\;\;\;\;
\;\;\;.\nonumber
\end{eqnarray}
Here we have used the fact that $\cos\varphi=\sin\varepsilon 
=O(\varepsilon)$ and that $|l_x|$ is largest for $x=0$ with 
$|l_x(x=0)|=\cot\varphi=\tan\varepsilon$ so that $l_x=O(\varepsilon)$. 
This leads to Eq.(3.7). From a systematic point of view in the last 
line of Eq.(B1) the 
prefactor $\sin\varphi=\cos\varepsilon$ can be dropped and $\cot^2
\varphi$ can be replaced by $\varepsilon ^2$. However, it 
turns out that it is advantageous to keep the full form of 
these terms (see Eqs.(3.9) and (3.10)).\\
\newpage

\centerline{\bf {Referrences}}
\begin{description}
\item{[1]  } {D. Andelman, J.F. Joanny, and 
M. O. Robbins, Europhys. Lett. {\bf 7}, 731 
(1988); M. O. Robbins, D. Andelman, J. F. Joanny, Phys. Rev. A {\bf 43}, 4344 
(1991); J. L. Harden and D. Andelman, Langmuir {\bf 8}, 2547 (1992).}
\item{[2]  } {P. Pfeifer, Y.J. Wu, M.W. Cole, and J. Krim, Phys. Rev. Lett. 
{\bf 62}, 1997 (1989); P. Pfeifer, M.W. Cole, and J. Krim, Phys. 
Rev. Lett. {\bf 65}, 663 (1990).}
\item{[3]  } {M. Kardar and J.O. Indekeu, Phys. Rev. Lett. {\bf 65}, 662 (1990); 
Europhys. Lett. {\bf 12}, 161 (1990); H. Li and
M. Kardar, Phys. Rev. B {\bf 42}, 6546 (1990); J. Krim and J.O. Indekeu, 
Phys. Rev. E {\bf 48}, 1576 (1993).}
\item{[4]  } {E. Cheng, M.W. Cole, and A.L. Stella, Europhys. Lett. {\bf 8}, 537 
(1989); E. Cheng, M.W. Cole, and P. Pfeifer, Phys. Rev. B {\bf 39}, 12962 
(1989).}
\item{[5]  } {G. Giugliarelli and A.L. Stella, Phys. Scripta T {\bf 35}, 34 
(1991); 
Phys. Rev. E {\bf 53}, 5035 (1996); Physica A {\bf 239}, 467 (1997); 
G. Sartoni, A.L. Stella, G. Giugliarelli, and M.R. D'Orsogna,  
Europhys. Lett. {\bf 39}, 633 (1997).}
\item{[6]  } {G. Palasantzas, Phys. Rev. B {\bf 48}, 14472 (1993); 
Phys. Rev. B {\bf 51}, 14612 (1995).}
\item{[7]  } {C. Borgs, J. De Coninck, R. Koteck\'y, and M. Zinque, 
Phys. Rev. Lett. {\bf 74}, 2293 (1995); K. Topolski, D. Urban, 
S. Brandon, and 
J. De Coninck, Phys. Rev. E {\bf 56}, 3353 (1997).}
\item{[8]  } {J.Z. Tang and J.G. Harris, J. Chem. Phys. {\bf 103}, 8201 (1995).}
\item{[9]  } {R. Netz and D. Andelman, Phys. Rev. E {\bf 55}, 687 (1997).}
\item{[10]} {A.O. Parry, P.S. Swain, and J.A. Fox, J. Phys.: Condens.
Matter {\bf 8}, L659 (1996); P.S. Swain and A.O. Parry, J. Phys. A: 
Math. Gen. {\bf 30}, 4597 (1997).}
\item{[11]} {D. Beaglehole, J. Phys. Chem. {\bf 93}, 893 (1989).}
\item{[12]} {S. Garoff, E.B. Sirota, S.K. Sinha, and H.B. Stanley, 
J. Chem. Phys. {\bf 90}, 7505 (1989).}
\item{[13]} {I. M. Tidswell, T.A. Rabedeau, P.S. Pershan, and S.D. Kosowsky, 
Phys. Rev. Lett. {\bf 66}, 2108 (1991), P.S. Pershan, Ber. Bunsenges. Phys. 
Chem. {\bf 98}, 372 (1994).}
\item{[14]} {V. Panella and J. Krim, Phys. Rev. E {\bf 49}, 4179 (1994).}
\item{[15]} {S. Dietrich, in {\it Phase Transitions and Critical Phenomena}, 
edited 
by C. Domb and J.L. Lebowitz (Academic, London, 1988), Vol.12, p. 1.}
\item{[16]} {M. Schick, in {\it Liquids at Interfaces,  Proceedings of the 
Les Houches Summer School in Theoretical Physics, Session XLVIII}, 
edited by J. Chavrolin, J. F. Joanny, and J. Zinn-Justin (North-Holland, 
Amsterdam, 1990), p. 415.}
\item{[17]} {W. Hansen, J.P. Kotthaus, and U. Merkt, in {\it Semiconductors and 
Semimetals}, edited by M. Reed (Academic, London, 1992), Vol. 35, p. 279.}
\item{[18]} {P. B\"{o}nsch, D. W\"{u}llner, T. Schrimpf, A. Schlachetzki, and 
R. Lacmann, J. Electrochem. Soc. {\bf 145}, 1273 (1998).}
\item{[19]} {D.W.L. Tolfree, Rep. Prog. Phys. {\bf 61}, 313 (1998).}
\item{[20]} {J.A. Mann, Jr., L. Romero, R.R. Rye, and F.G. Yost, 
Phys. Rev. E {\bf 52}, 3967 (1995); R.R. Rye, F.G. Yost, and J.A. Mann, 
Langmuir {\bf 12}, 555 (1996); {\it ibid} 4625 (1996).}
\item{[21]} {S. Gerdes and G. Str\"{o}m, Colloids and Surfaces A {\bf 116}, 135 
(1996); S. Gerdes, A.-M. Cazabat, and G. Str\"{o}m, Langmuir {\bf 13}, 7258 
(1997).}
\item{[22]} {M. Dong, F.A. Dullien, and I. Chatzis, J. Coll. Interf. Sci. 
{\bf 172}, 21 (1995); M. Dong and I. Chatzis, J. Coll. Interf. Sci. 
{\bf 172}, 278 (1995); D. Zhou, M. Blunt, and F.M. Orr, Jr., J. Coll. 
Interf. Sci. {\bf 187}, 11 (1997).}
\item{[23]} {E. Kim and G.M. Whitesides, J. Phys. Chem. B {\bf 101}, 855 
(1997).}
\item{[24]} {J.B. Knight, A. Vishwanath, J. P. Brody, and R.H. Austin,  Phys. 
Rev. Lett. {\bf 80}, 3863 (1998).}
\item{[25]} {M. Trau, N. Yao, E. Kim, Y. Xia, G.M. Whitesides, and I.A. Aksay, 
Nature {\bf 390}, 674 (1997).}
\item{[26]} {T.A. Winningham, H.P. Gillis, D.A. Choutov, K.P. Martin, 
J. T. Moore, and K. Douglas, Surf. Sci. {\bf 406}, 221 (1998).}
\item{[27]} {F. Burmeister, C. Sch\"{a}fle, B. Keilhofer, C. Bechinger, 
J. Boneberg, and P. Leiderer, Adv. Mater. {\bf 10}, 495 (1998).}
\item{[28]} {B.V. Derjaguin and N.V. Churaev, J. Coll. Inter. Sci. {\bf 54}, 157 
(1976).}
\item{[29]} {J.R. Philip, J. Chem. Phys. {\bf 66}, 5069 (1977); {\it ibid} 
{\bf 67}, 1732 (1977).}
\item{[30]} {Y. Pomeau, J. Coll. Interf. Sci. {\bf 113}, 5 (1985).}
\item{[31]} {E. Cheng and M.W. Cole, Phys. Rev. B {\bf 41}, 9650 (1990).}
\item{[32]} {P.M. Duxbury and A.C. Orrick, Phys. Rev. B {\bf 39}, 2944 (1989).}
\item{[33]} {M. Napi\'orkowski, W. Koch, and S. Dietrich, Phys. Rev. A {\bf 45}, 
5760 (1992).}
\item{[34]} {E.H. Hauge, Phys. Rev. A {\bf 46}, 4944 (1992).}
\item{[35]} {G.A. Darbellay and J. Yeomans, J. Phys. A: Math. Gen. {\bf 25}, 
4275 (1992).}
\item{[36]} {E. Cheng, J.R. Banavar, M.W. Cole, and F. Toigo, Surf. Sci. {\bf 
261}, 
389 (1992).}
\item{[37]} {Y. Song, Phys. Lett. A {\bf 180}, 3611 (1993).}
\item{[38]} {A. Koroci\'nski and M. Napi\'orkowski, Mol. Phys. {\bf 84}, 
171 (1995).}
\item{[39]} {M. Kagan, W. V. Pinczewski, and P. E. Oren, J. Coll. Interf. Sci. 
{\bf 170}, 426 (1995).}
\item{[40]} {M. Schoen and S. Dietrich, Phys. Rev. E {\bf 56}, 499 (1997).}
\item{[41]} {D. Henderson, S. Soko{\l}owski, and D. Wasan, Phys. Rev E {\bf 57}, 
5539 (1998).}
\item{[42]} {A. Lipowski, Phys. Rev. E {\bf 58}, R1 (1998).}
\item{[43]} {P. M\"{u}ller-Buschbaum, M. Tolan, W. Press, F. Brinkop, and J.P. 
Kotthaus, Ber. Bunsenges. Phys. Chem. {\bf 98}, 413 (1994).}
\item{[44]} {Z. Li, M. Tolan, T. H\"{o}hr, D. Kharas, S. Qu, J. Sokolov, 
M.P. Rafailovich, H. Lorenz, J.P. Kotthaus, J. Wang, S.K. Sinha, and A. Gibaud, 
Macromolecules {\bf 31}, 1915 (1998).}
\item{[45]} {For an overview of adsorption of fluids on structured substrates 
see 
S. Dietrich, in Proceedings of the NATO-ASI {\it New Approaches to Old and New 
Problems in Liquid State Theory - Inhomogeneities and Phase Separation in 
Simple,  Complex and Quantum Fluids} held at Patti Marina (Messina), Italy, 
July 7-17, 1998, edited by C. Caccamo (Kluwer, Dordrecht), in press.}
\item{[46]} {I.M. Gelfand and S.V. Fomin, {\it Calculus of Variations} 
(Prentice-Hall, Englewood Cliffs, 1963), Sec. 14.}
\item{[47]} {T. Onda, S. Shibuichi, N. Satoh, and K. Tsujii, Langmuir {\bf 12}, 
2125 (1996); S. Shibuichi, T. Onda, N. Satoh, and K. Tsujii, J. Phys. Chem. 
{\bf 100}, 19512 (1996).}
\item{[48]} {J.K. Lee and H.I. Avronson, Surf. Sci. {\bf 47}, 692 (1975).}
\item{[49]} {R.K.P. Zia, J.E. Avron, and J.E. Taylor, J. Stat. Phys. {\bf 50}, 
727 
(1988).}
\item{[50]} {D. Langbein, J. Fluid Mech. {\bf 213}, 251 (1990).}
\item{[51]} {J. De Coninck, J. Fruttero, and A. Ziermann, Physica A {\bf 196}, 
320 
(1993); {\it ibid} {\bf 199}, 243  (1993).}
\item{[52]} {L.-H. Tang and Y. Tang, J. Phys. II France {\bf 4}, 881 (1994).}
\item{[53]} {P. Concus and R. Finn, Phys. Fluids {\bf 10}, 39 (1998).}
\item{[54]} {S. Dietrich, in {\it Phase Transitions in Surface Films 2}, 
Proceedings of the NATO-ASI (Series B) held in Erice, Italy, 19-30 June 1990, 
edited by H. Taub, G. Torzo, H.J. Lauter, and S.C. Fain, Vol. B {\bf 267}, 
391 (1991).}
\item{[55]} {S. Dietrich and M. Napi\'orkowski, Phys. Rev. A {\bf 43}, 1861 
(1991).}
\item{[56]} {M. Kagan and W.V. Pinczewski, J. Coll. Interf. Sci. {\bf 203}, 
379 (1998).}
\item{[57]} {J. O. Indekeu, Physica A {\bf 183}, 439 (1992); Int. J. Mod. Phys. 
B {\bf 8}, 309 (1994).}
\item{[58]} {S. Dietrich and M. Napi\'orkowski, Physica A {\bf 177}, 437 (1991); 
M. Napi\'orkowski and S. Dietrich, Z. Phys. B {\bf 89}, 263 (1992); Phys. Rev. E 
{\bf 57}, 1836 (1993); Z. Phys. B {\bf 97}, 511 (1995).} 
\item{[59]} {M.E. Fisher and A.J. Jin, Phys. Rev. B {\bf 44}, 1430 (1991); 
Phys. Rev. Lett. {\bf 69}, 792 (1992); A.J. Jin and M.E. Fisher, 
Phys. Rev. B {\bf 47}, 7365 (1993); {\it ibid}  {\bf 48}, 2642 (1993).}
\item{[60]} {E.H. Hauge and M. Schick, Phys. Rev. B {\bf 27}, 4288 (1983); M.P. 
Nightingale, W.F. Saam, and M. Schick, Phys. Rev. B {\bf 30}, 3830 (1984).}
\item{[61]} {V. Privman and M.E. Fisher, J. Stat. Phys. {\bf 33}, 385 (1983); J. 
Appl. Phys. {\bf 57}, 3327 (1985).}
\item{[62]} {T. Bieker and S. Dietrich, Physica A {\bf 252}, 85 (1998); due to a 
printer's error in Eq.(4.6) 
of this reference ${\sum}_{s}$ should read ${\sum}_{l}$.}
\item{[63]} {$I(y)=-{Li}_{2}(-y)$ can be expressed in terms of the dilogarithmic 
function ${Li}_{2}(z)=-\int_{0}^{z}\,dz\,z^{-1}\,ln(1-z)$. See L. Levin, 
{\it Polylogarithms and Associated Functions} (North Holland, New York, 
1981) and L. Levin (ed.), in {\it Mathematical Surveys and Monographs}, Vol. 37, 
{\it Structural Properties of Polylogarithms} (American Mathematical 
Society, Providence, 1991).}
\item{[64]} {H.W. Diehl, in {\it Phase Transitions and Critical Phenomena}, 
edited 
by C. Domb and J.L. Lebowitz (Academic, London, 1986), Vol. 10, p. 76.}
\item{[65]} {R. Lipowsky, Z. Phys. B {\bf 55}, 345, (1984).}
\item{[66]} {K. Rejmer and M. Napi\'orkowski, Z. Phys. B {\bf 97}, 293 (1995).}
\item{[67]} {K. Rejmer and M. Napi\'orkowski, Z. Phys. B {\bf 102}, 101 (1997).}
\item{[68]} {M.E. Fisher, A.J. Jin, and A.O. Parry, Ber. Bunsenges. Phys. Chem. 
{\bf 98}, 357 (1994).}
\end{description}    
\newpage

\centerline{\bf Figure captions}

\noindent
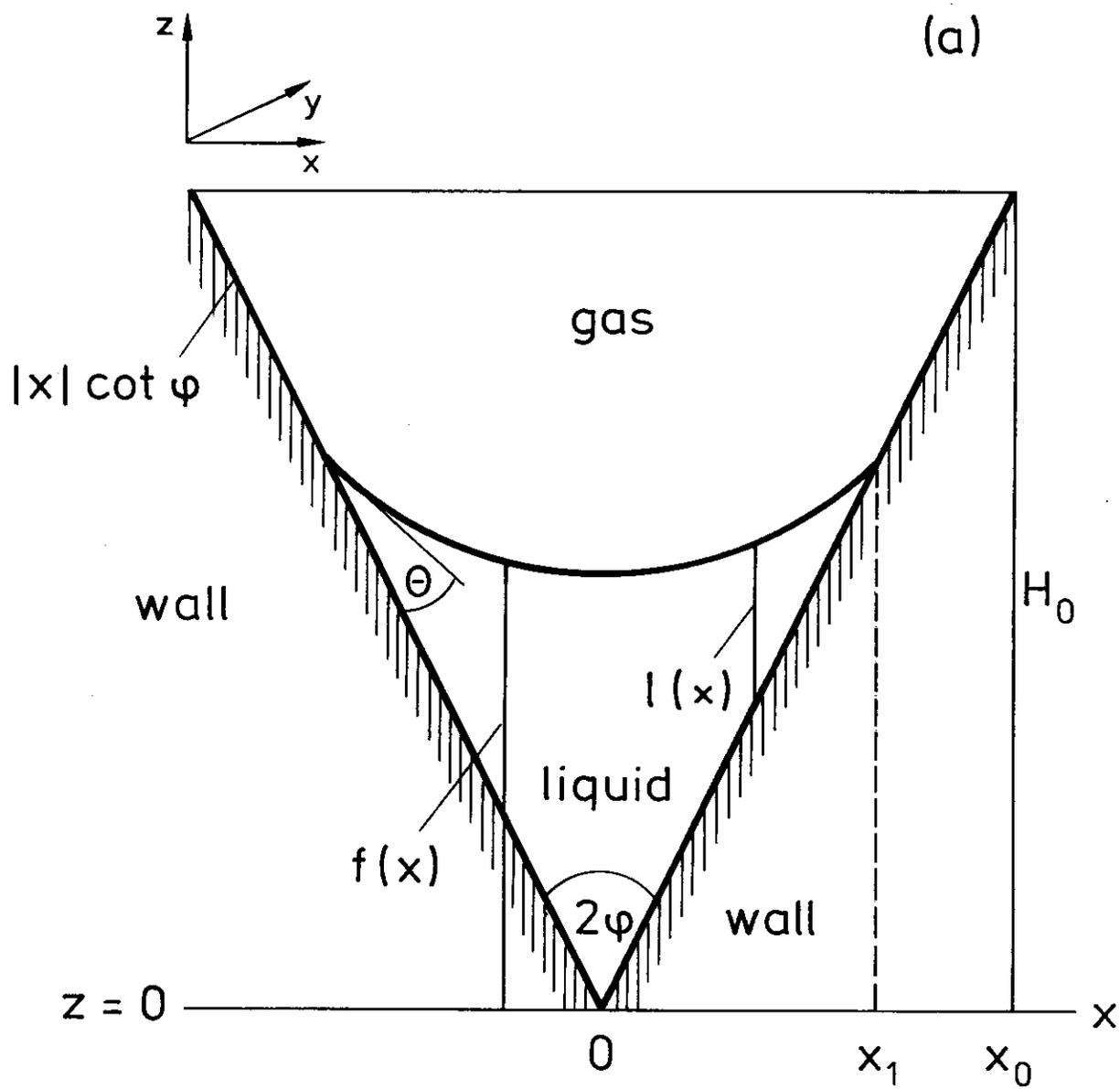
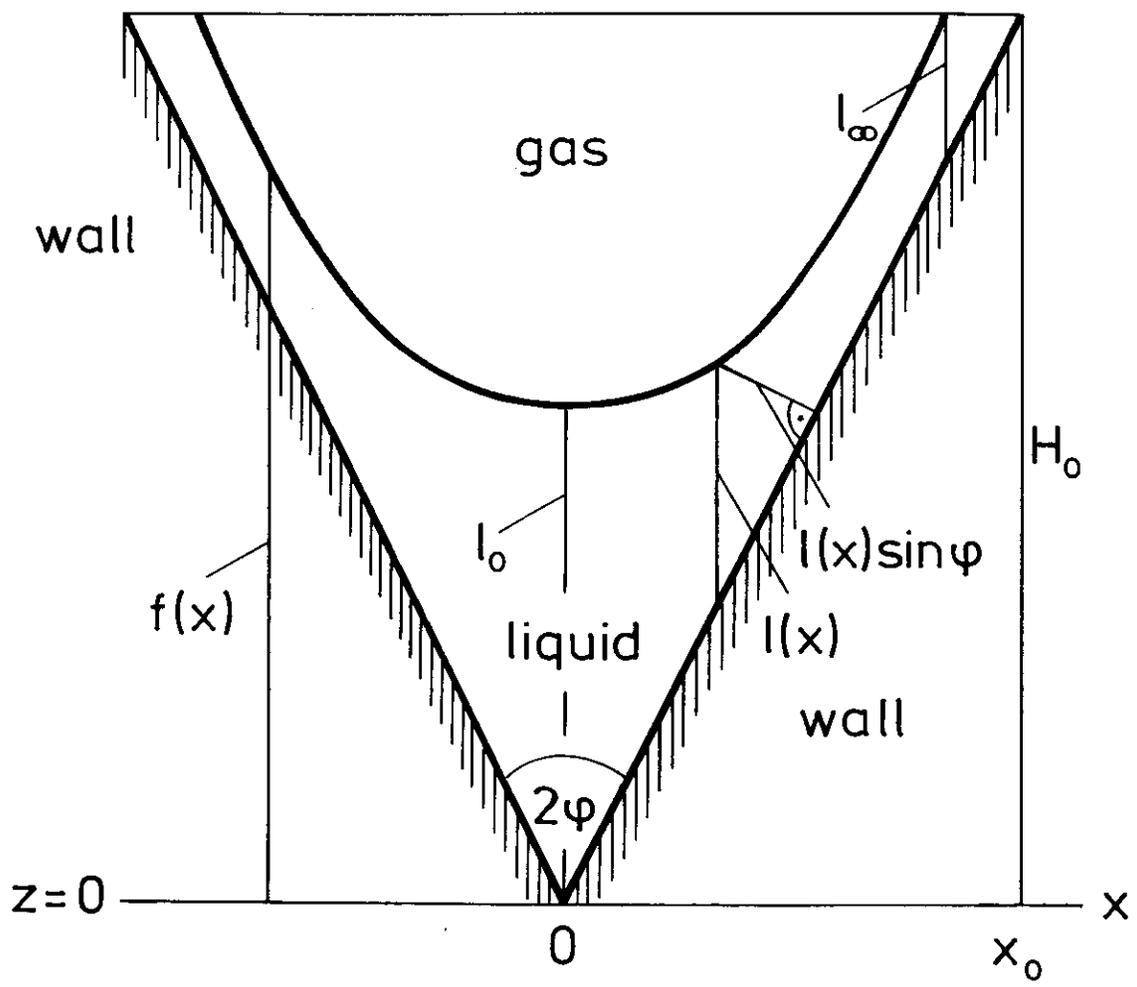
FIG.1.(a) Macroscopic description of a wedge with opening angle
$2\,\varphi$ formed by identical walls whose surfaces are located
at $z=|x|\cot\varphi$. The shape of the meniscus is
described by $z=f(x)$ or $l(x)=f(x)-|x|\cot\varphi$.
The liquid-gas interface intersects the walls at $x=\pm x_1$
with a contact angle $\Theta$. The system is taken 
to be two-dimensional and its height is cut off at $z=H_0$.
(b) Same as in (a) on a microscopic scale which takes into account
that far away from the center of the wedge the meniscus reduces to
a wetting film of thickness $l_{\infty}\sin\varphi$ covering
the walls at large $|x|$; $l_{\infty}=l(x\rightarrow\infty)$.
$l_0\equiv l(x=0)$ denotes the filling height of the liquid in 
the wedge.\\

\noindent
FIG.2. Classification of the equilibrium shapes of a nonvolatile
liquid drop (shaded area) as function of the opening angle $2\varphi$ of
a symmetric wedge  and of the contact angle $\Theta$. In all cases
the shape of the liquid-vapor interface is a part of a circle. The space
within the wedge, which is not filled by the liquid, is occupied by
vapor. Here the wedge is two-dimensional resembling the situation
in which a three-dimensional system exhibits translational invariance
along the edge of the wedge.\\

\noindent
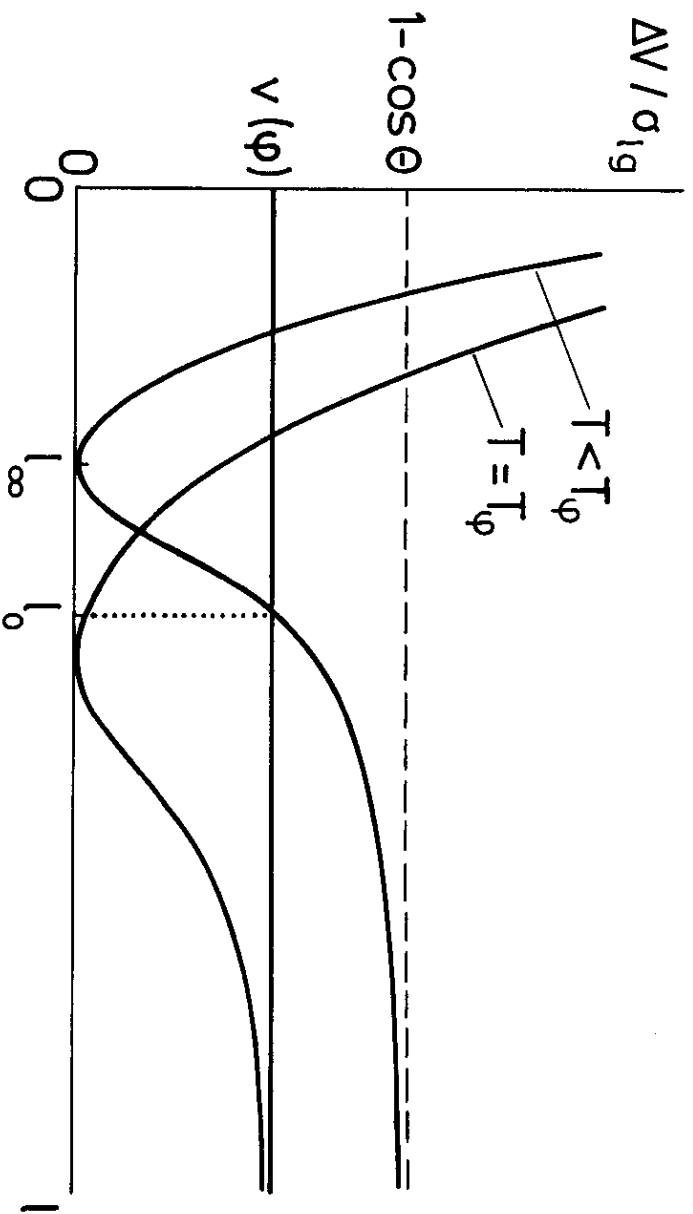
FIG.3. Schematic graphical solution for the filling height $l_0$
according to Eqs.(3.5) and (3.10), respectively, for the case of
a critical wetting transition of the corresponding planar substrate,
i.e., $\Delta\mu=0$. In this case $\Delta V$ exhibits a single
minimum at $l=l_{\infty}$ which moves smoothly to infinity
for $T\rightarrow T_w$ and becomes more shallow for increasing 
temperature. For $l\rightarrow \infty$ the ratio 
$\Delta V/\sigma_{lg}$ attains the limiting value
$1-\cos\Theta(T)$ which vanishes for $T\rightarrow T_w$.
By construction $\Delta V$ is positive. The intersection with
the straight line $v(\varphi)$ yields the filling height $l_0$.
There are two solutions $l_0^{(1)}$ and $l_0^{(2)}$ but only the solution
$l_0>l_{\infty}$ is compatible with the boundary condition
associated with Eq.(3.9) (see Subsec.IIIC). For $T\nearrow T_{\varphi}$
the asymptote $1-\cos\Theta(T)$ approaches $v(\varphi)$
from above so that $l_0$ diverges continuously 
for $T\nearrow T_{\varphi}$.\\

\noindent
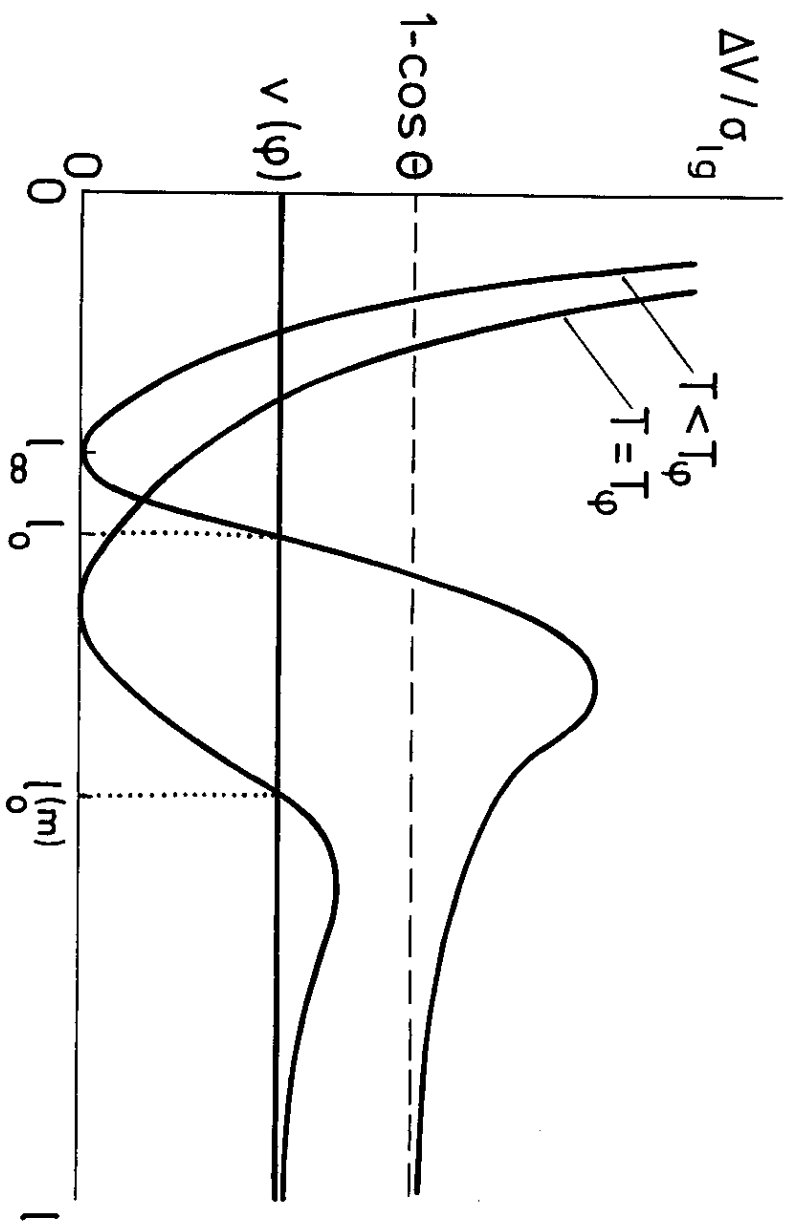
FIG.4. Same as Fig.3 for the case of a first-order wetting of
the corresponding planar substrate. For $T\nearrow T_{\varphi}$
the difference
$1-\cos\Theta(T)$ reaches $v(\varphi)$ and $l_0$ increases smoothly
to a finite maximum value $l_0^{(m)}$ at $T=T_{\varphi}$.
At $T_{\varphi}$ there is another solution $l_0=\infty$. As will
be shown in Subsec.IIIC, $l_0=\infty$ is the thermodynamically 
stable solution for $T>T_{\varphi}$. Therefore at $T_{\varphi}$
the filling height
$l_0$ jumps from the finite value $l_0^{(m)}$ to infinity.\\

\noindent
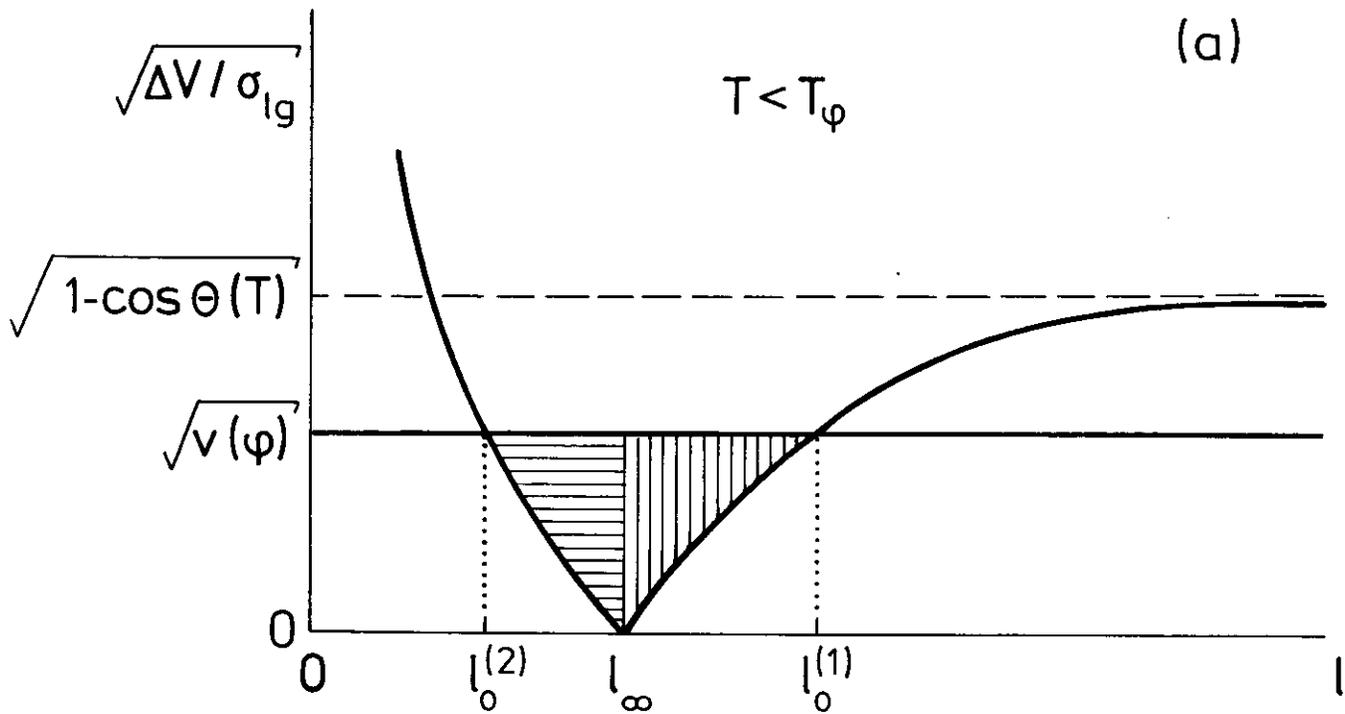
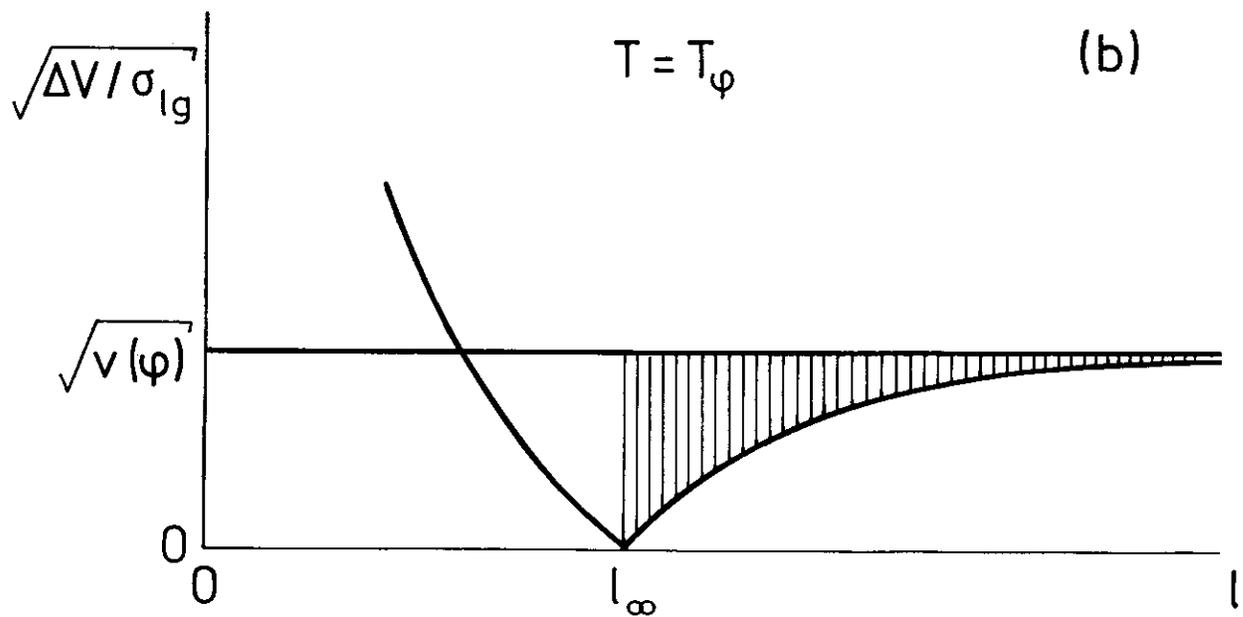
FIG.5. Schematic graphical interpretation of the line tension
in the case of critical wetting of the corresponding planar
substrate below (a) and at (b) the filling transition temperature.
According to Eqs.(3.13) and (3.14), for $l_0=l_0^{(1)}$
the line tension $\eta$ is proportional to the vertically hatched area $A$,
which is to be taken negatively, whereas for $l=l_0^{(2)}$ the line tension
$\eta$ is proportional to the horizontally hatched area, which 
is to be taken positively. Therefore $l_0^{(1)}$ has the lower
line tension and is thermodynamically stable.
For $T\nearrow T_{\varphi}$ $l_{\infty}$ increases, the minimum of
$\sqrt{\Delta V}$ becomes less steep, and $l_0^{(1)}$ moves out to infinity
corresponding to the filling of the wedge. Even for dispersion
forces with $V(L\rightarrow\infty)\sim L^{-2}$ the area $A$
and the line tension $\eta$ remain finite for  $T\nearrow T_{\varphi}$.
The second value of $l_0$, i.e., $l_0^{(2)}$ does
not correspond to solution of Eq.(3.4) or (3.9).\\

\noindent
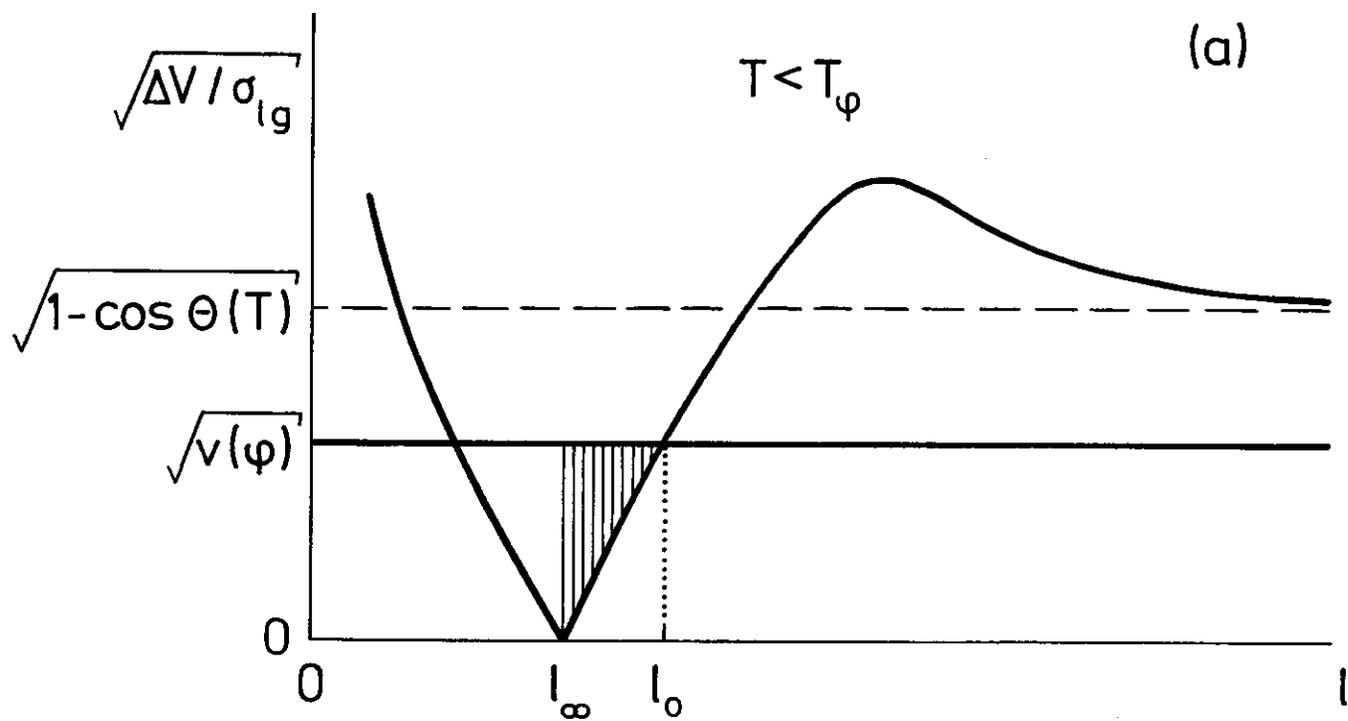
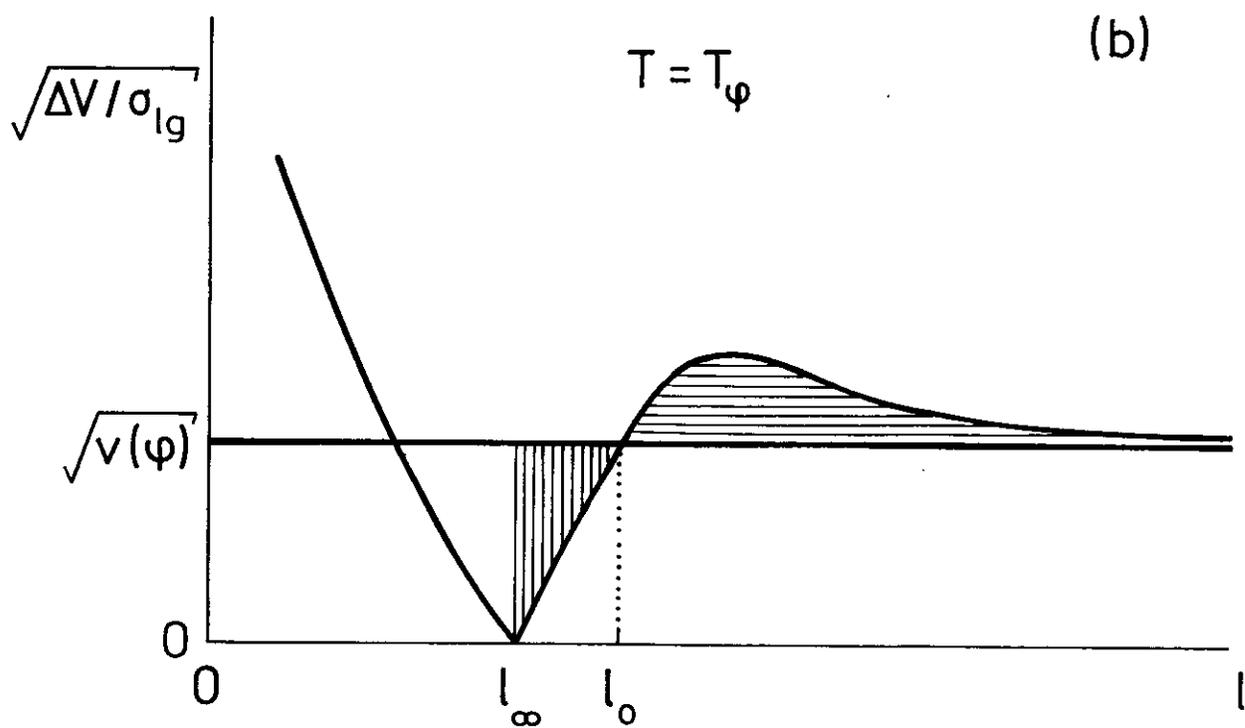
FIG.6. Same as Fig.5 for first-order wetting of the corresponding
planar substrate. For  $T\nearrow T_{\varphi}$ both $l_{\infty}$
and $l_0$ increase but stay finite. At $T_{\varphi}$ 
the filling height $l_0=\infty$ 
is also a solution, but the corresponding line tension is larger 
than that corresponding to the indicated finite solution by
an amount given by  the
horizontally hatched area, which is to be taken positively. For
$T>T_{\varphi}$ the surface free energy favors the filled wedge
so that $l_0(T)$ undergoes a discontinuous jump from a finite value at
$T=T_{\varphi}^{-}$ to a macroscopically large value at  
$T_{\varphi}^{+}$.\\

\noindent
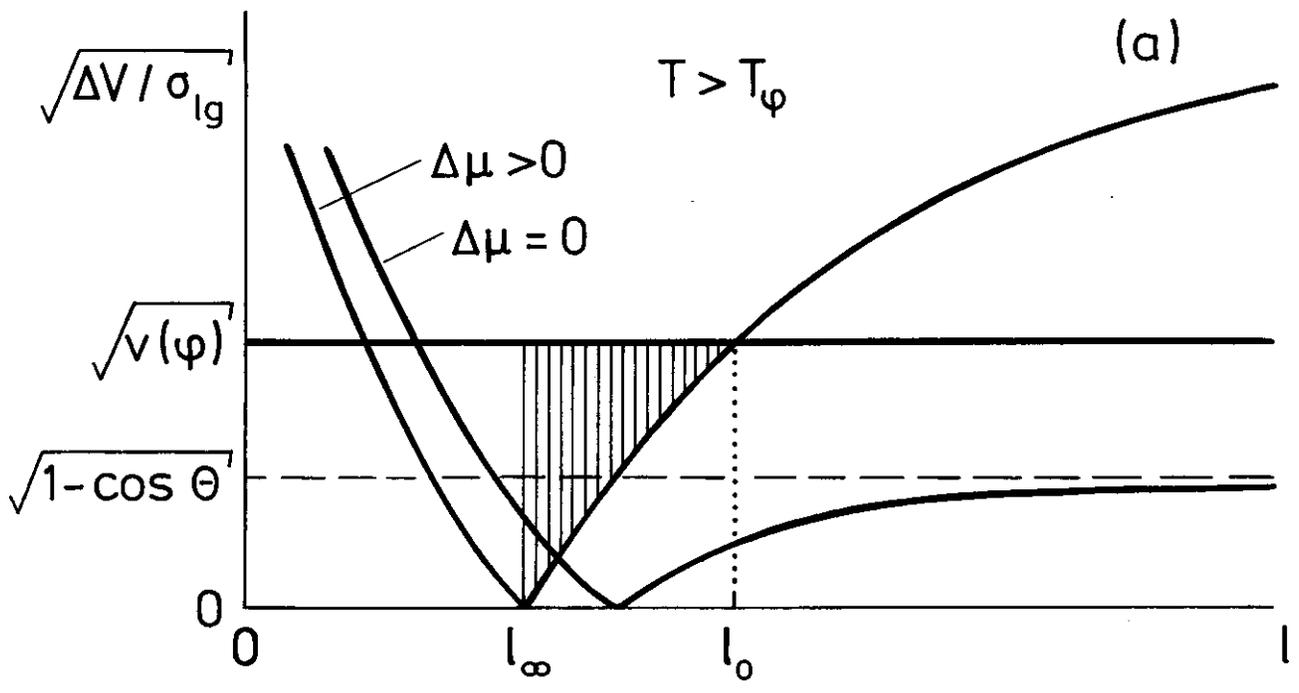
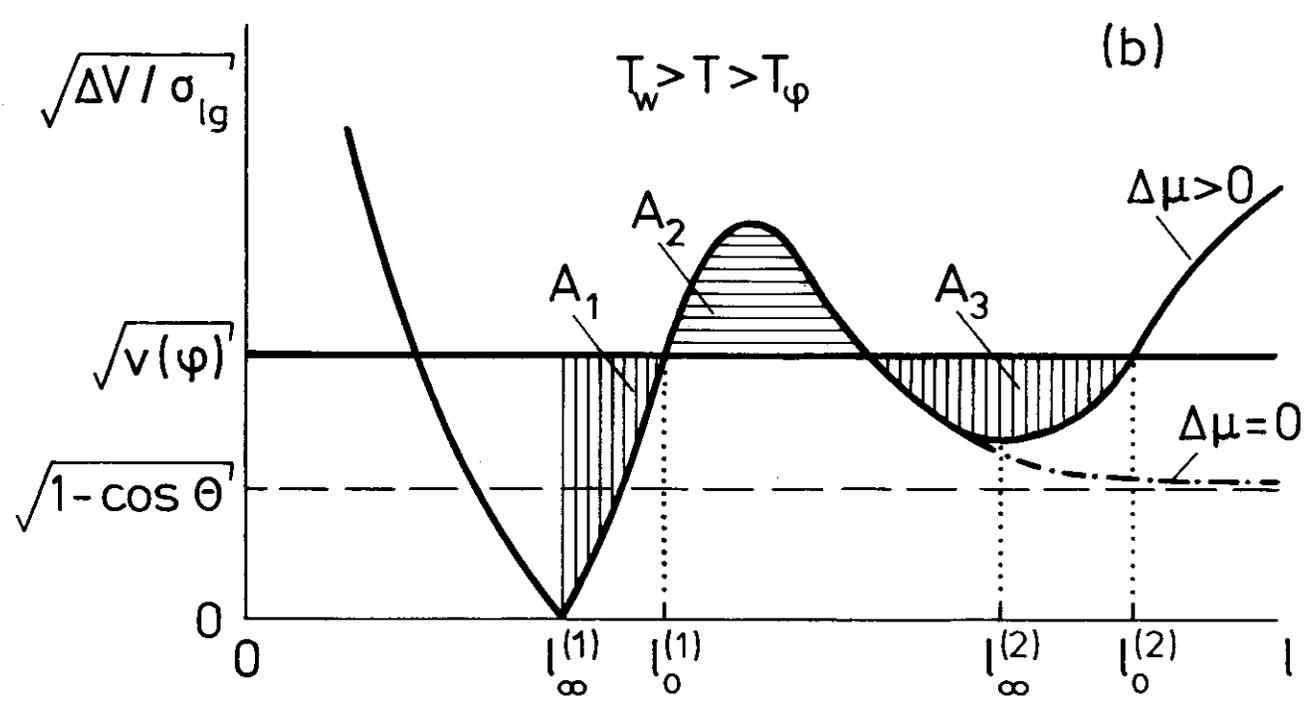
FIG.7. Schematic illustration of the filling transition of the wedge
along isotherms $\Delta \mu\rightarrow 0$ above the filling 
temperature $T_{\varphi}$ for critical (a) and first-order wetting
(b) of the corresponding planar substrate.
Off coexistence $\sqrt{\Delta V/\sigma_{lg}}$ increases for large 
$l$ as $(\Delta\mu\Delta\rho\sin\varphi/\sigma_{lg})
^{\frac{1}{2}}l^{\frac{1}{2}}$ whereas $\sqrt{\Delta V/\sigma_{lg}}$
attains the finite value $\sqrt{1-\cos\Theta}$ for $\Delta\mu=0$
which is less than $\sqrt{v(\varphi)}$ for $T>T_{\varphi}$.
In the case of critical wetting there is a single filling height $l_0$
which can possibly be thermodynamically stable. 
For $\Delta\mu\rightarrow 0$ the wetting
film thickness $l_{\infty}$ increases slightly but
remains finite whereas $l_0$ diverges continuously. 
The vertically hatched area diverges also so that
$\eta(\Delta \mu\rightarrow 0,T>T_{\varphi})\rightarrow -\infty$.
In the case of first-order wetting $\sqrt{\Delta V/\sigma_{lg}}$
exhibits a global minimum at $l_{\infty}^{(1)}$ and a local minimum at
$l_{\infty}^{(2)}$. Upon crossing the prewetting line of the corresponding 
planar substrate $l_{\infty}^{(2)}$ turns into the global minimum. Thus (b)
corresponds to a case $T_w>T>T_{\varphi}$. For large $\Delta\mu$
the area $|A_3|=-A_3$ is smaller then $A_2>0$ so
that $A_2+A_3>0$. Consequently $l_0^{(1)}$ corresponds to
the global minimum of the line tension. 
For $\Delta\mu\rightarrow 0$ the area
$|A_3|$ increases without limit so that there is a critical value
$\Delta\mu_{pf}(T)$ at which $|A_3|=A_2$ so that for 
$\Delta\mu<\Delta\mu_{pf}(T)$ the filling heigth $l_{0}^{(2)}$ becomes 
the globally stable configuration. Upon lowering $\Delta\mu$
further $l_0^{(2)}$ 
diverges, as well as $A_1+A_2+A_3$, leading to the filling
of the wedge. For reasons of clarity we have ignored the slight changes in
$\sqrt{\Delta V/\sigma_{lg}}$ for $l\leq l_{\infty}^{(2)}$
upon lowering $\Delta\mu$. $\Delta\mu_{pf}(T)$ marks a prefilling
transition in the wedge.\\

\noindent
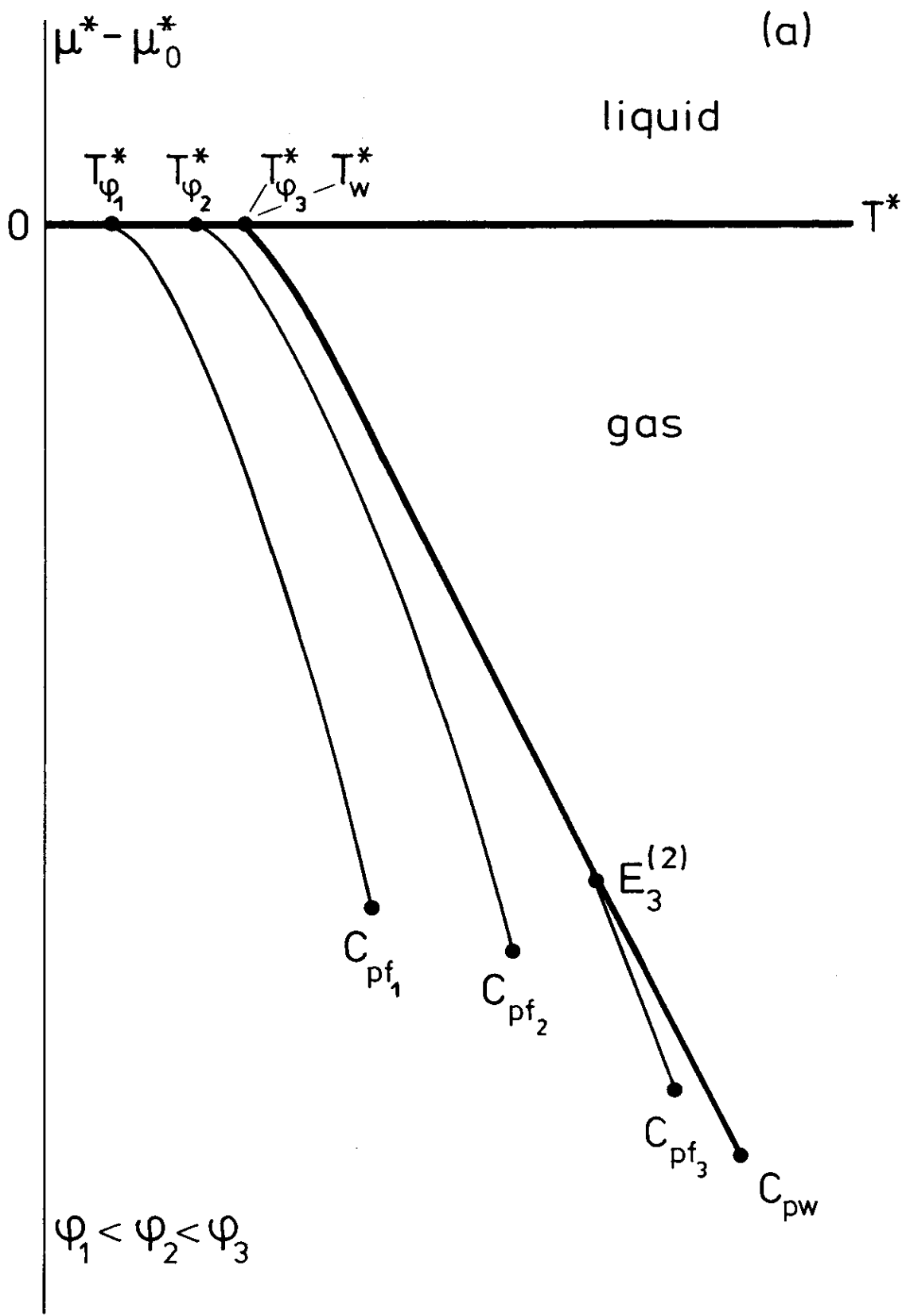
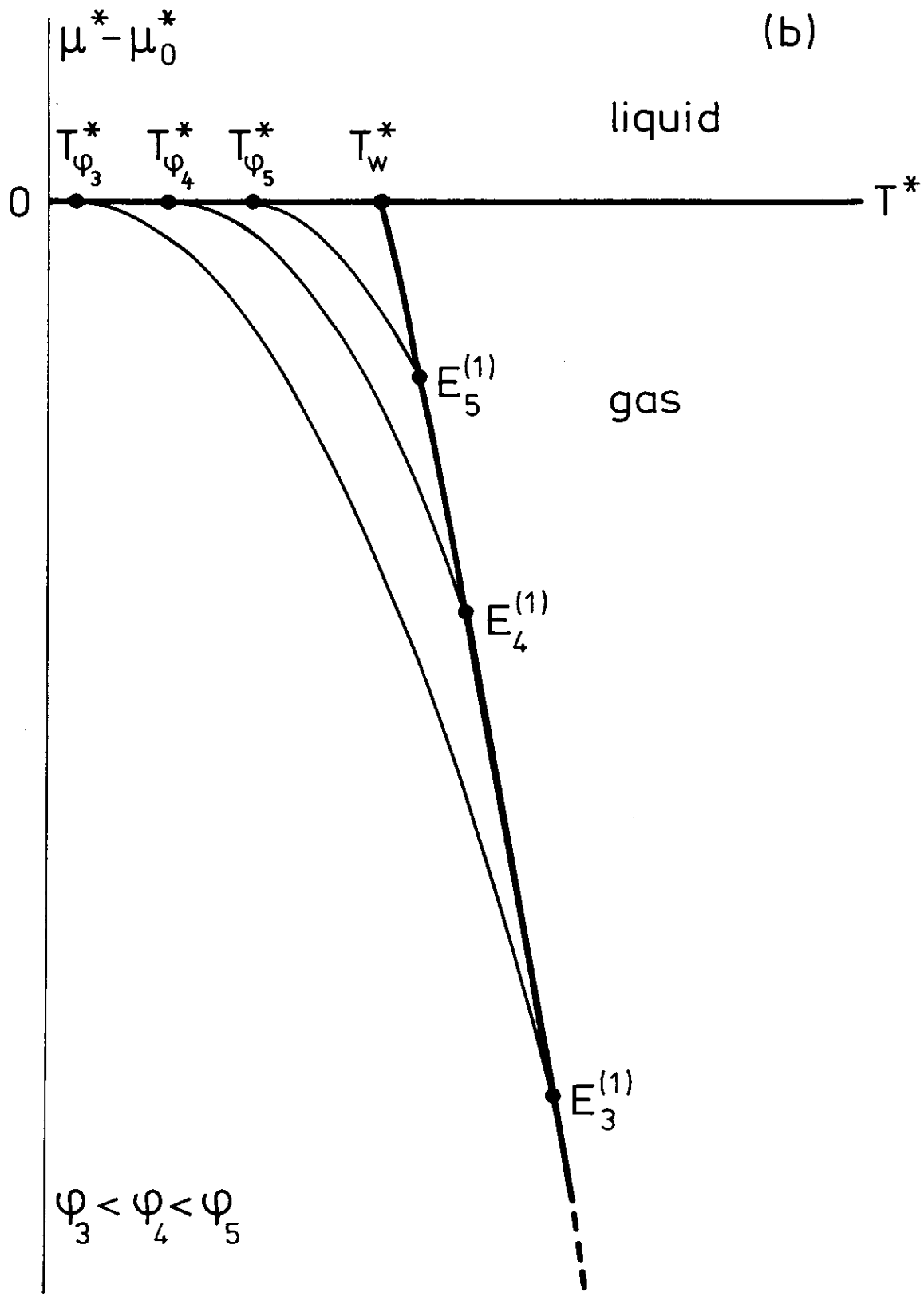
FIG.8. (a) Phase diagram for the filling of a wedge in the case
that the corresponding planar substrate exhibits a first-order
wetting transition at $T_w$.
The thick phase boundaries represent the gas-liquid coexistence
curve at $\mu=\mu_0$, which for reasons of simplicity has been
taken to be a straight line, and the prewetting line emanating
from $T_w$ and ending at a prewetting critical point $C_{pw}$.
The bulk critical point $T_c$ is off the scale to the right.
We use dimensionless quantities $\mu^{*}-\mu_0^{*}
=-\Delta\mu\Delta\rho\xi /W_w$ (see the main text)
and $T^{*}=T/T_w$ so that $T_w^{*}\,=\,1$. The temperatures
$T_{\varphi_{i}}$ denote 
first-order filling transition temperatures which move towards 
$T_w$ for increasing values of $\varphi_{i}$. From each filling 
transition point a so-called prefilling line emanates ending in a 
critical prefilling point $C_{pf}$. The prefilling lines 
join the gas-liquid coexistence line tangentially as a quadratic function
whereas the corresponding tangential approach of the prewetting line
is logarithmic and not visible on the present scale.
For $\varphi\rightarrow\frac{1}{2}\pi$ the prefilling line touches
the prewetting line and breaks into two pieces. For $\varphi\,=\,
\varphi_3$ in (a) the lower piece between the extraordinary point
$E_3^{(2)}$ and $C_{pf}$ is shown. On this scale the upper piece between
$T_{\varphi_{3}}$ and $E_3^{(1)}$ is not visible
as well as the distinction between $T_w$ and $T_{\varphi_3}$. 
This is resolved in (b) on a magnified scale. These phase diagrams 
have been obtained for the model defined by Eqs.(3.7), (3.10),
(3.2), and (3.17) using the following potential parameters:
$C/W_w^2=3.504823$, and $U/W_w=0.197338$. For the angles
$\varphi_1=81.40^{o}$, $\varphi_2=83.12^{o}$, $\varphi_3=84.84^{o}$, 
$\varphi_4=85.99^{o}$, and $\varphi_5=87.13^{o}$  
with $W_{{\varphi}_i} = W(T_{{\varphi}_i})$ one obtains 
$W_{\varphi_1}/W_w=1.008580$, $W_{\varphi_2}/W_w=1.005532$,
$W_{\varphi_3}/W_w=1.003130$, $W_{\varphi_4}/W_w=1.001899$,
and, $W_{\varphi_5}/W_w=1.000971$ so that 
$T_{w}^{*}\,-\,T_{{\varphi}_{i}}^{*}\,=\,\frac{W_{w}}{W_{0}}
\left(\frac{W_{{\varphi}_{i}}}{W_{w}}-1\right)$. Putting numbers on the 
axis requires to choose a value for the ratio $\frac{W_{w}}{W_{0}}$.  
For $\frac{W_{w}}{W_{0}}=1$ one obtains  
$T_{w}^{*}\,-\,T_{{\varphi}_{i}}^{*}\,=\,0.008580, 0.005532, 0.003130, 
0.001899, 0.000971$,  for i=1,...,5 and $T^{*}_{C_{pw}}=1.594132$ and 
${\mu}^{*}(T_{C_{pw}})\,-\,{\mu}^{*}_{0}\,=\,-0.038024$. For a different 
value of $\frac{W_{w}}{W_{0}}$ the temperatures are rescaled linearly 
according to the formula given above. 
\\

\noindent
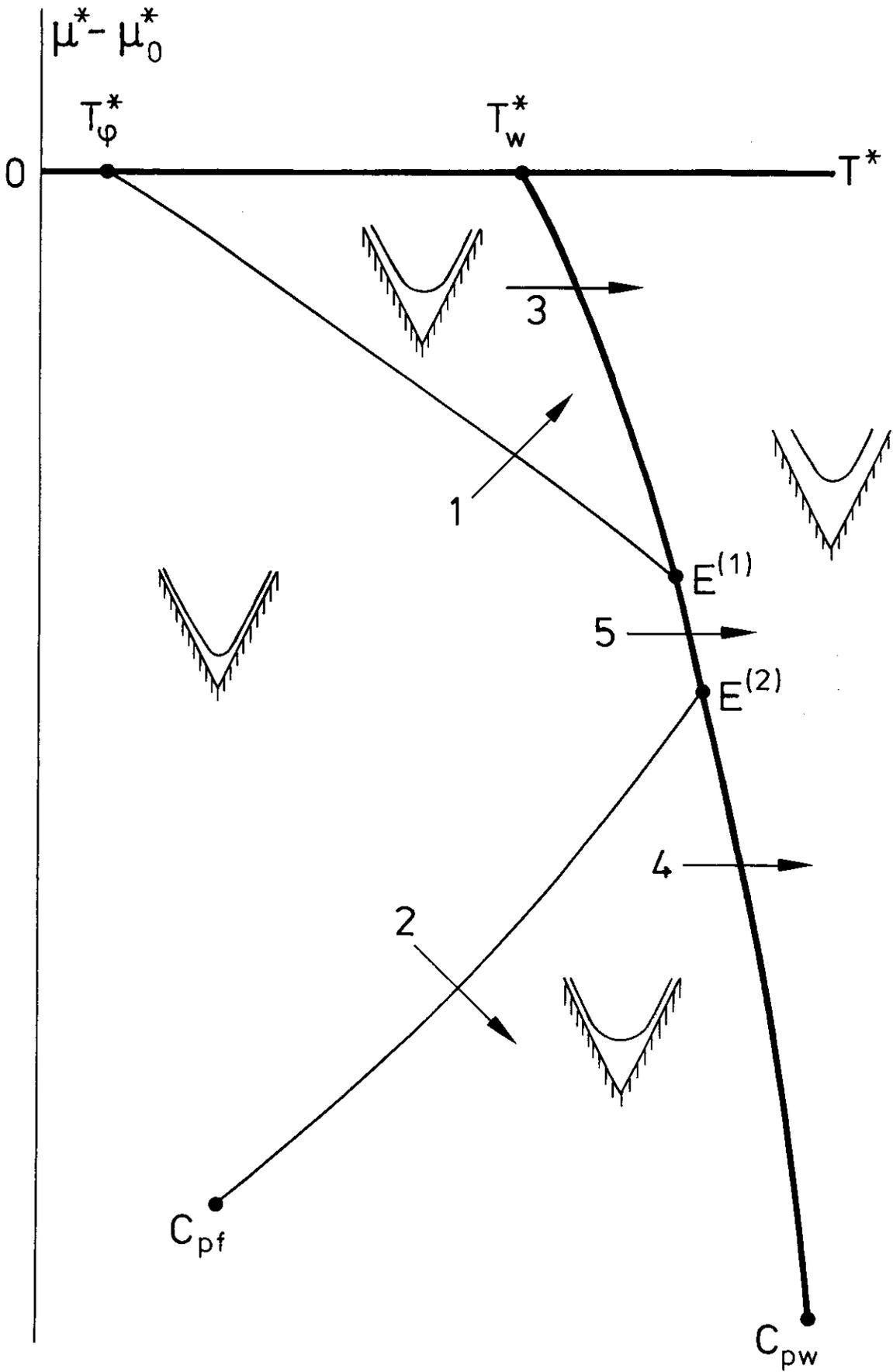
FIG.9. Types of morphology of the wetting film in a wedge for the  
various phases within a schematic phase diagram. 
The notation is the same as in Fig.8. The opening
angle $\varphi$ is sufficiently large so that the prefilling line
is split into two pieces forming the two extraordinary points
$E^{(1)}$ and $E^{(2)}$. Along the thermodynamic paths $1$ and $2$
the filling height $l_0$ in the center of the wedge increases
discontinuously upon crossing the pieces of the prefilling line
but the thickness $l_{\infty}$ of the wetting film far away from 
the center of the wedge does not jump. Along the thermodynamic
paths $3$ and $4$ $l_{\infty}$ increases discontinuously while
$l_0$ grows smoothly upon crossing the prewetting line. Along
the thermodynamic path $5$ both $l_0$ and $l_{\infty}$ jump only
at the prewetting line.\\

\noindent
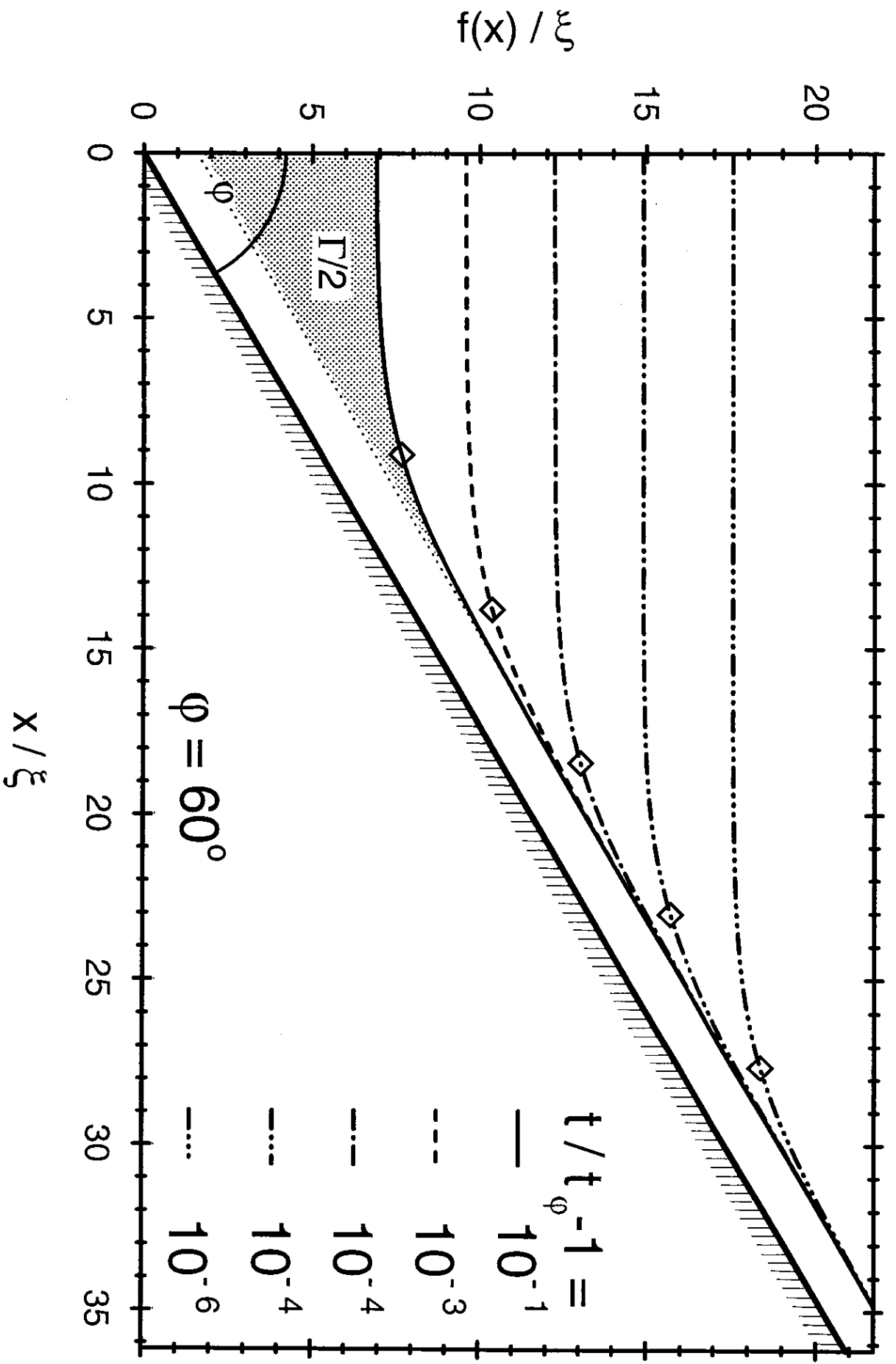
FIG.10. Shape $f(x)=f(-x)$ of the meniscus in units of the bulk
correlation length $\xi$ in the liquid phase. $f(x > 0)=l(x)+x
\cot\varphi$ with $l(x)$ given by Eqs.(4.2) and (4.9). The planar
substrate undergoes a critical wetting transition at $T_w$.
The temperature is raised towards the filling transition temperature
$T_{\varphi}$, i.e., $t/t_{\varphi}=(T_w-T)/(T_w-T_{\varphi})
\rightarrow 1$ along two-phase coexistence in the bulk.
In units of $\xi$ and in terms of $t/t_{\varphi}$ the profile
is determined uniquely  by the dimensionless parameter
$\sqrt{2U/\sigma_{lg}}$ which is chosen to be $2$ here.
The diamonds indicate the position of the maximum curvature
of $f(x)$. As explained in the main text  this position
attains  a constant distance from the wall for $t/t_{\varphi}
\rightarrow 1$. The shaded area corresponds to half of the 
wedge excess coverage $\Gamma/\Delta\rho$ (see Eq.(4.13))
for $t/t_{\varphi}=10^{-2}$. The dotted line indicates 
the asymptote of $f(x)$ extended to $x=0$. At the present scale
the temperature dependence of the asymptotes, i.e., of
$l_{\infty}$ is not visible.\\

\noindent
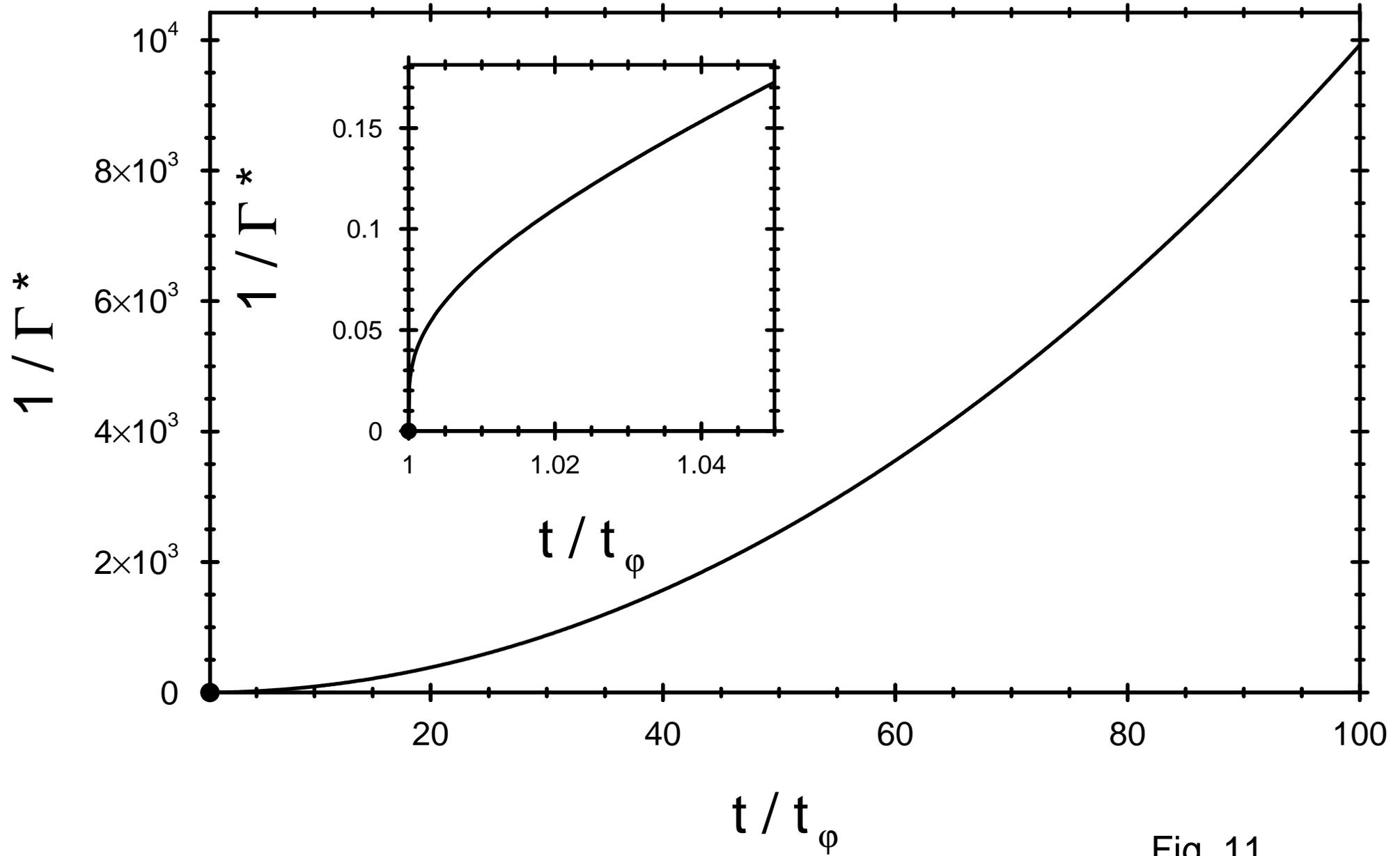
FIG.11. Inverse of the reduced excess coverage
$\Gamma ^{*}=\Gamma(\sin(2\varphi))/(4\Delta\rho\xi^2)$
as function of reduced temperature $t/t_{\varphi}$
(Eq.(4.13)). Upon approaching the filling transition temperature
$\Gamma^{*}$ diverges $\sim \ln^2(t/t_{\varphi}-1)$ 
(see Eq.(4.14) and the inset). For $\varphi\rightarrow \frac{1}{2}\pi$, 
i.e., $t/t_{\varphi}\rightarrow\infty$
the inverse coverage $1/\Gamma ^{*}$ diverges quadratically
so that $\Gamma \sim \Gamma ^{*} /\sin (2\varphi) \sim
\varphi - \frac{1}{2}\pi$ for $\varphi \rightarrow \frac{1}{2}\pi$,
i.e, $t/t_{\varphi} \rightarrow \infty$.
Note that in terms of the variable $t/t_{\varphi}$ the functional
form of $\Gamma^{*}$ is independent of $\varphi$ and the model parameters 
$\sigma_{lg}$, $W_{0}$, and $U$.\\

\noindent
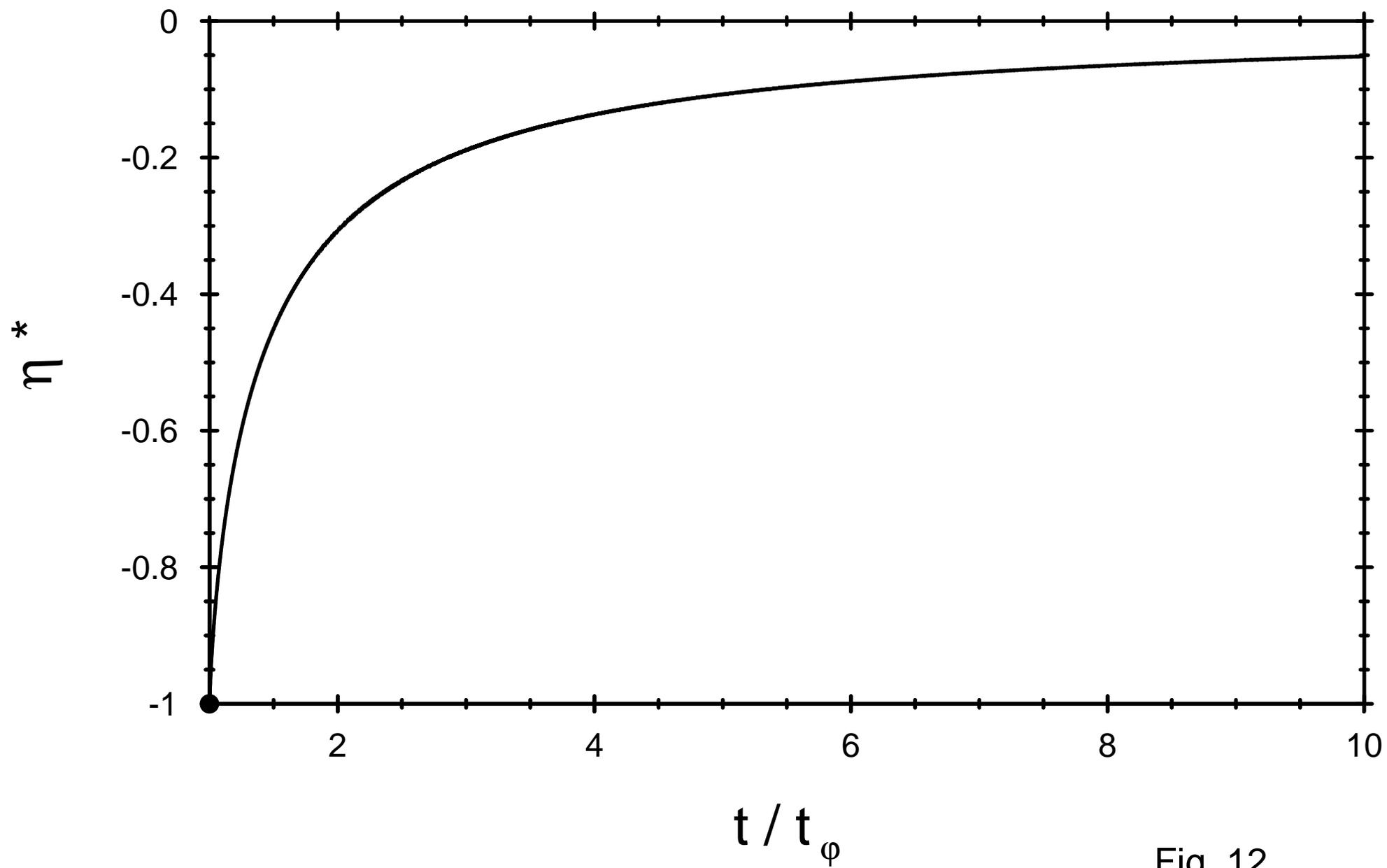
FIG.12. Reduced line tension $\eta ^{*}=\eta/(2\xi\sigma_{lg}
\cot\varphi)$ as function of $t/t_{\varphi}$ (Eq.(4.15)).
$\eta^{*}$ is negative and attains its minimum at $t=t_{\varphi}$
in a cusplike singularity $\sim (t/t_{\varphi}-1)\ln(t/t_{\varphi}-1)$
(Eq.(4.16)). For $\varphi \rightarrow \frac{1}{2}\pi$, i.e., 
$t/t_{\varphi}\rightarrow \infty$, the line tension $\eta^{*}$
vanishes as $-\frac{1}{2}t/t_{\varphi}$ so that $\eta 
\sim \eta^{*}\cot\varphi \sim (\varphi-\frac{1}{2}\pi)^2$
for $\varphi \rightarrow \frac{1}{2}\pi$. In terms of the variable 
$t/t_{\varphi}$ the functional form of $\eta^{*}$ is independent
of $\varphi$ and the model parameters $\sigma_{lg}$, $W_{0}$, and $U$.
 
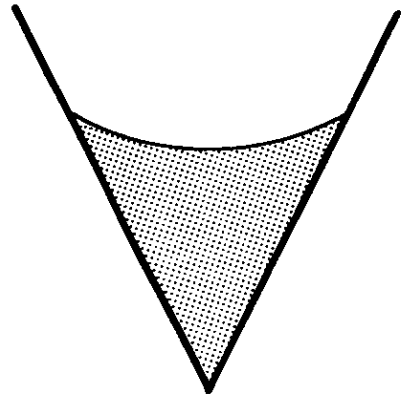
\end{document}